\documentclass[fleqn,usenatbib]{mnras}
\usepackage{newtxtext,newtxmath}
\usepackage[T1]{fontenc}
\usepackage{ae,aecompl}

\usepackage{graphicx}	
\usepackage{color}
\usepackage{amsmath}	
\usepackage{amssymb}	
\usepackage{booktabs}
\usepackage{longtable}

\title[The Metal-Rich Atmosphere of HAT-P-26b]{The Metal-Rich Atmosphere of the Neptune HAT-P-26b}

\author[MacDonald \& Madhusudhan]{
Ryan J. MacDonald$^{1}$\thanks{Email: r.macdonald@ast.cam.ac.uk}
\& Nikku Madhusudhan$^{1}$\thanks{Email: nmadhu@ast.cam.ac.uk}
\\
$^{1}$Institute of Astronomy, University of Cambridge, Madingley Road, Cambridge, CB3 0HA, UK
}

\date{Accepted 13 March 2019. Received 8 March 2019; in original form 23 January 2019}

\pubyear{2019}

\begin{document}
\label{firstpage}
\pagerange{\pageref{firstpage}--\pageref{lastpage}}
\maketitle

\begin{abstract}
Transmission spectroscopy is enabling precise measurements of atmospheric H$_2$O abundances for numerous giant exoplanets. For hot Jupiters, relating H$_2$O abundances to metallicities provides a powerful probe of their formation conditions. However, metallicity measurements for Neptune-mass exoplanets are only now becoming viable. Exo-Neptunes are expected to possess super-solar metallicities from accretion of H$_2$O-rich and solid-rich planetesimals. However, initial investigations into the exo-Neptune HAT-P-26b suggested a significantly lower metallicity than predicted by the core-accretion theory of planetary formation and solar system expectations from Uranus and Neptune. Here, we report an extensive atmospheric retrieval analysis of HAT-P-26b, combining all available observations, to reveal its composition, temperature structure, and cloud properties. Our analysis reveals an atmosphere containing $1.5^{+2.1}_{-0.9}\%$ H$_2$O, an O/H of $18.1^{+25.9}_{-11.3} \, \times$ solar, and C/O $< 0.33$ (to 2$\sigma$). This updated metallicity, the most precise exo-Neptune metallicity reported to date, suggests a formation history with significant planetesimal accretion, albeit below that of Uranus and Neptune. We additionally report evidence for metal hydrides at 4.1$\sigma$ confidence. Potential candidates are identified as TiH (3.6$\sigma$), CrH (2.1$\sigma$), or ScH (1.8$\sigma$). Maintaining gas-phase metal hydrides at the derived temperature ($563^{+58}_{-54}$ K) necessitates strong disequilibrium processes or external replenishment. Finally, we simulate the \emph{James Webb Space Telescope} Guaranteed Time Observations for HAT-P-26b. Assuming a composition consistent with current observations, we predict JWST can detect H$_2$O (at 29$\sigma$), CH$_4$ (6.2$\sigma$), CO$_2$ (13$\sigma$), and CO (3.7$\sigma$), thereby improving metallicity and C/O precision to 0.2 dex and 0.35 dex, respectively. Furthermore, NIRISS observations could detect several metal hydrides at $>$ 5$\sigma$ confidence.

\end{abstract}


\section{Introduction}\label{Intro}
Transmission spectroscopy has opened an unprecedented window into the atmospheric composition of exoplanets. Recent years have seen detections of multiple atomic and molecular species \citep[e.g][]{Snellen2010,Deming2013,Macintosh2015,Sedaghati2017,Hoeijmakers2018}. Beyond detections, atmospheric retrieval techniques have enabled measurements of the abundances of chemical species \citep[see][for a recent review]{Madhusudhan2018}. By constraining the atmospheric composition, especially the H$_2$O inventory, crucial clues are provided to identify formation scenarios \citep{Oberg2011,Madhusudhan2014b,Mordasini2016}. The most precise constraints have arisen from studies of hot Jupiters, as their extended atmospheres enhance the viability of transmission spectroscopy. Studies of hot Jupiters have reported H$_2$O abundances implying a range of O/H ratios, from nearly solar ($\sim 0.05\%$) \citep{Kreidberg2014,Line2016a,Sing2016} to significantly sub-solar \citep{Madhusudhan2014a,Barstow2016,Pinhas2019}.

By contrast, the atmospheric composition of lower mass exoplanets, including exo-Neptunes and super-Earths, has proven challenging to constrain. In the solar system, lower mass planets possess atmospheres with an increasing fraction of heavy elements -- termed the atmospheric \emph{metallicity}. The metallicity of the solar system giants is commonly expressed in terms of the atmospheric C/H ratio, as derived from CH$_4$ abundances. For Jupiter and Saturn, C/H is $\sim 4 \times$ solar and $\sim 10 \times$ solar, respectively \citep{Atreya2016}, whilst the ice-giants Uranus and Neptune are $\sim 80 \times$ solar \citep{Karkoschka2011,Sromovsky2011}. This trend is consistent with the core-accretion theory of planet formation \citep{Pollack1996}. If low mass exo-Neptunes form in the same manner, they are anticipated to contain substantially H$_2$O enriched atmospheres due to accretion of water-rich planetesimals \citep{Fortney2013}. Alternatively, in situ formation close to the parent star, resulting in minimal contamination by planetesimals, should lead to H$_2$/He dominated atmospheres similar to the stellar photosphere \citep{Rogers2011}. H$_2$O abundances thereby offer insights into the accretion history and physical properties of the original planetesimal building blocks. Measuring the composition of exo-Neptunes thus provides a powerful avenue to differentiate between planet formation mechanisms.

Until recently, constraints on exo-Neptune metallicities have proven relatively inconclusive. GJ 436b possesses a flat transmission spectrum, frustrating attempts to measure its composition and detect H$_2$O. One scenario is GJ 436b's atmosphere is dominated by high-altitude clouds \citep{Knutson2014}, with another possibility being a high metallicity due to accretion of rocky planetesimals, as suggested by its dayside emission spectrum \citep{Madhusudhan2011,Moses2013,Morley2017}. The first detection of H$_2$O in an exo-Neptune was reported by \citet{Fraine2014} for HAT-P-11b's atmosphere, though the derived metallicity (1-700$\times$ solar, to $3 \sigma$) is consistent with both a nearly pure H$_2$/He envelope and a wide range of core accretion scenarios.

Recently, the transmission spectrum of the exo-Neptune HAT-P-26b has provided a high-significance H$_2$O detection \citep{Wakeford2017}. HAT-P-26b is a 4.23 day period exoplanet with $M_p = 18.6 M_{\earth}$ and $R_p = 6.33 R_{\earth}$ \citep{Hartman2011,Wakeford2017}, implying $g_p = 4.47$ ms$^{-2}$. The combination of low gravity and high temperature ($T_{\mathrm{eq}}$ = 990 K) results in an extended atmosphere ideal for transmission spectroscopy. The first observations of HAT-P-26b's transmission spectrum, using \emph{Magellan} and \emph{Spitzer}, were obtained by \citet{Stevenson2016}. They reported evidence of H$_2$O, though their data could not differentiate between a high metallicity ($\sim 100 \times$ solar) clear atmosphere and a solar metallicity atmosphere with a 10 mbar cloud deck. \citet{Wakeford2017} obtained additional visible and infrared transmission spectra with \emph{Hubble}. This spectrum enabled them to report the first well-constrained metallicity in an exo-Neptune: O/H = $4.8^{+21.5}_{-4.0} \times$ solar (to $1 \sigma$) -- notably smaller than the $\sim 60 \times$ solar that would be expected for a planet of this mass from core-accretion scenarios \citep{Fortney2013}.

However, no attempt has been made to simultaneously deduce the implications of these two datasets. In this work, we present a comprehensive atmospheric retrieval analysis, using all available observations, to derive the composition, metallicity, and other properties of HAT-P-26b's atmosphere. As the lowest mass exoplanet with a detected spectral feature in primary transit, HAT-P-26b is presently our best window into formation mechanisms of Neptune-mass exoplanets. HAT-P-26b will also be observed during the \emph{James Webb Space Telescope}'s (JWST) Guaranteed Time Observations (GTO), for which we offer predictions.

In what follows, the observations of HAT-P-26b are outlined in \S\ref{section:observations}. We describe our atmospheric models and retrieval architecture in \S\ref{section:methods}. The derived atmospheric properties of HAT-P-26b are presented in \S\ref{section:results}. Predictions for JWST's GTO observations are offered in \S\ref{section:JWST}. Finally, in \S\ref{section:discussion} we summarise our results and discuss the implications.

\section{The Transmission Spectrum of HAT-P-26b}\label{section:observations}

\begin{figure}
	\includegraphics[width=\columnwidth,  trim={0.6cm 1.0cm 0.4cm 0.6cm}]{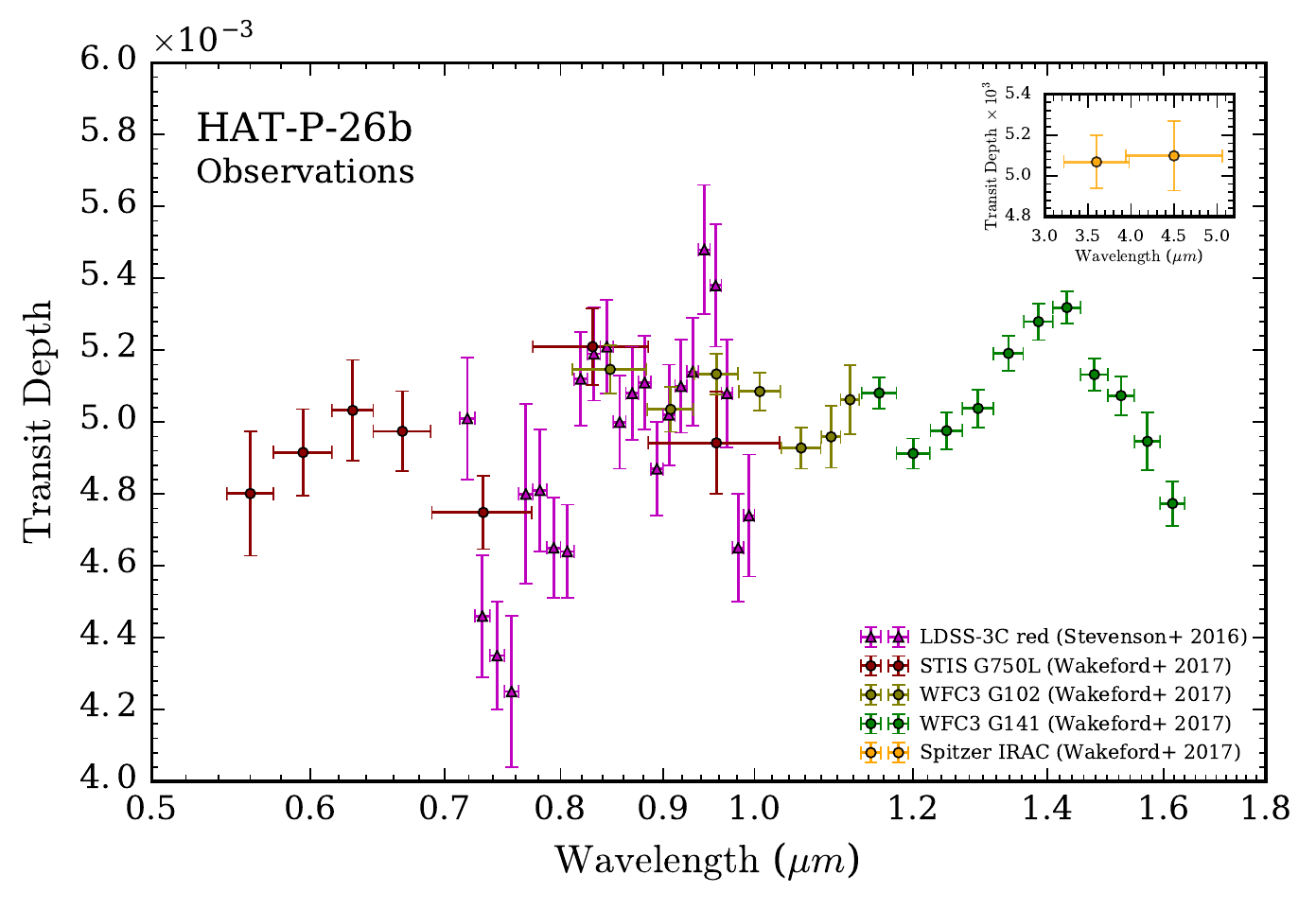}
    \caption{\textbf{Transit observations of HAT-P-26b}. Each transit depth marker is shaped according to source (triangles for \citet{Stevenson2016}, circles for \citet{Wakeford2017}) and coloured by instrument mode. Horizontal bars indicate bin widths, whilst vertical error bars correspond to 1-$\sigma$ precisions. \textbf{Inset}: Spitzer observations further in the infrared.}
    \label{fig:data}
\end{figure}

The transmission spectrum of HAT-P-26b comprises 50 observations from both ground and space-based facilities. \emph{Magellan} Low Dispersion Survey Spectrograph 3 (LDSS-3C) observations \citep{Stevenson2016} are complimented by \emph{Hubble} Space Telescope Imaging Spectrograph (STIS) and Wide Field Camera 3 (WFC3) grism observations, along with \emph{Spitzer} Infrared Array Camera (IRAC) photometric observations \citep{Wakeford2017}. The observed transit depths in each instrument mode are shown in Figure \ref{fig:data}.

The combined set of observations offers continuous wavelength coverage across the visible and near-infrared, from $0.5 - 1.6 \, \micron$. The Spitzer IRAC channels additionally cover the regions surrounding 3.6 \& 4.5 $\micron$. The STIS G750L observations span the range 0.5-1.0 $\micron$, with a mean spectral resolution $R \equiv \Delta \lambda/\lambda \approx 20$ and mean $(R_p/R_{*})^{2}$ precision $\sim 130$ ppm. The LDSS-3C red observations span the range 0.7-1.0 $\micron$, with mean $R \approx$ 70 and mean precision $\sim$ 155 ppm. The WFC3 G102 and G141 grism observations span the ranges 0.8-1.1 $\micron$ and 1.1-1.6 $\micron$, respectively, with corresponding resolutions $R \approx$ 40 and 60, and mean precisions of $\sim$ 70 and 50 ppm. The precision of the two Spitzer IRAC photometric observations are $\sim$ 130 and 170 ppm.

Our analysis considers these observations as a given input. The data reduction of the transit observations is discussed in detail in \citet{Stevenson2016,Wakeford2016,Sing2016,Wakeford2017}. We note that Spitzer observations are available from both \citet{Stevenson2016} and \citet{Wakeford2017}. Here we employ only those from \citet{Wakeford2017}, to ensure consistent data reduction across the space-based observations. To account for the possibility of differing normalisations between the ground-based and space-based observations due to stellar variability or differing reduction procedures, we allow for a relative offset between these two datasets during our analysis. With the observations described, we proceed to detail our framework for inverting these observations to obtain the atmospheric state of HAT-P-26b.

\section{Atmospheric Retrieval Architecture} \label{section:methods}

We infer the atmospheric properties at HAT-P-26b's day-night terminator using the radiative transfer and retrieval code POSEIDON \citep{MacDonald2017a}. This code couples a transmission spectrum forward model with a Bayesian parameter estimation and model comparison algorithm, enabling derivation of statistically rigorous atmospheric parameter constraints and detection significances for various atmospheric model components. POSEIDON has previously been applied to the atmospheric retrieval of hot Jupiters \citep{MacDonald2017a,MacDonald2017b,Sedaghati2017,Kilpatrick2018}. Here, we outline salient aspects as they apply to the present analysis - in particular, recent updates to generalise its scope to exo-Neptune atmospheres.

We describe our atmospheric models in section \ref{subsection:atmospheric_models}, generation of transmission spectra in section \ref{subsection:radiative_transfer}, and the parameters specifying each model atmosphere in section \ref{subsection:parameters}.

\subsection{Atmospheric models} \label{subsection:atmospheric_models}

We model the terminator region of HAT-P-26b's atmosphere in a plane-parallel geometry. The atmosphere is discretised into 100 axially-symmetric layers about the observer's line-of-sight, uniformly spaced in log-pressure from $10^{-7}$ to $10^{2}$ bar. The temperature in each layer, expressed as a function of pressure, determines the number density in each layer via the ideal gas law. The atmospheric composition, described in section \ref{subsubsection:chem_opac}, specifies the atmospheric mean molecular weight. The planetary radius and surface gravity are placed at a parameterised reference pressure, where they iteratively specify a radial distance grid under the assumption of hydrostatic equilibrium. Clouds and hazes, along with partial cloud coverage, are included as potential components of the atmosphere, as discussed in section \ref{subsubsection:cloud_opac}.

\subsubsection{Molecular \& atomic opacities} \label{subsubsection:chem_opac}

A wide range of potential chemical species and compositions need to be considered when modelling the atmospheres of exo-Neptunes \citep{Madhusudhan2016}. On the one-hand, atmospheres could be H$_2$-He dominated with other chemical species, such as H$_2$O and CO, present as trace gases -- as for hot Jupiters with near-solar or sub-solar metallicity. Alternatively, a wide range of high mean molecular weight atmospheres, especially H$_2$O or CO$_2$ rich compositions, are also a possibility at higher metallicities \citep{Moses2013}.

\begin{figure}
	\includegraphics[width=0.97\columnwidth, trim={0.6cm 0.4cm 0.6cm 0.4cm}]{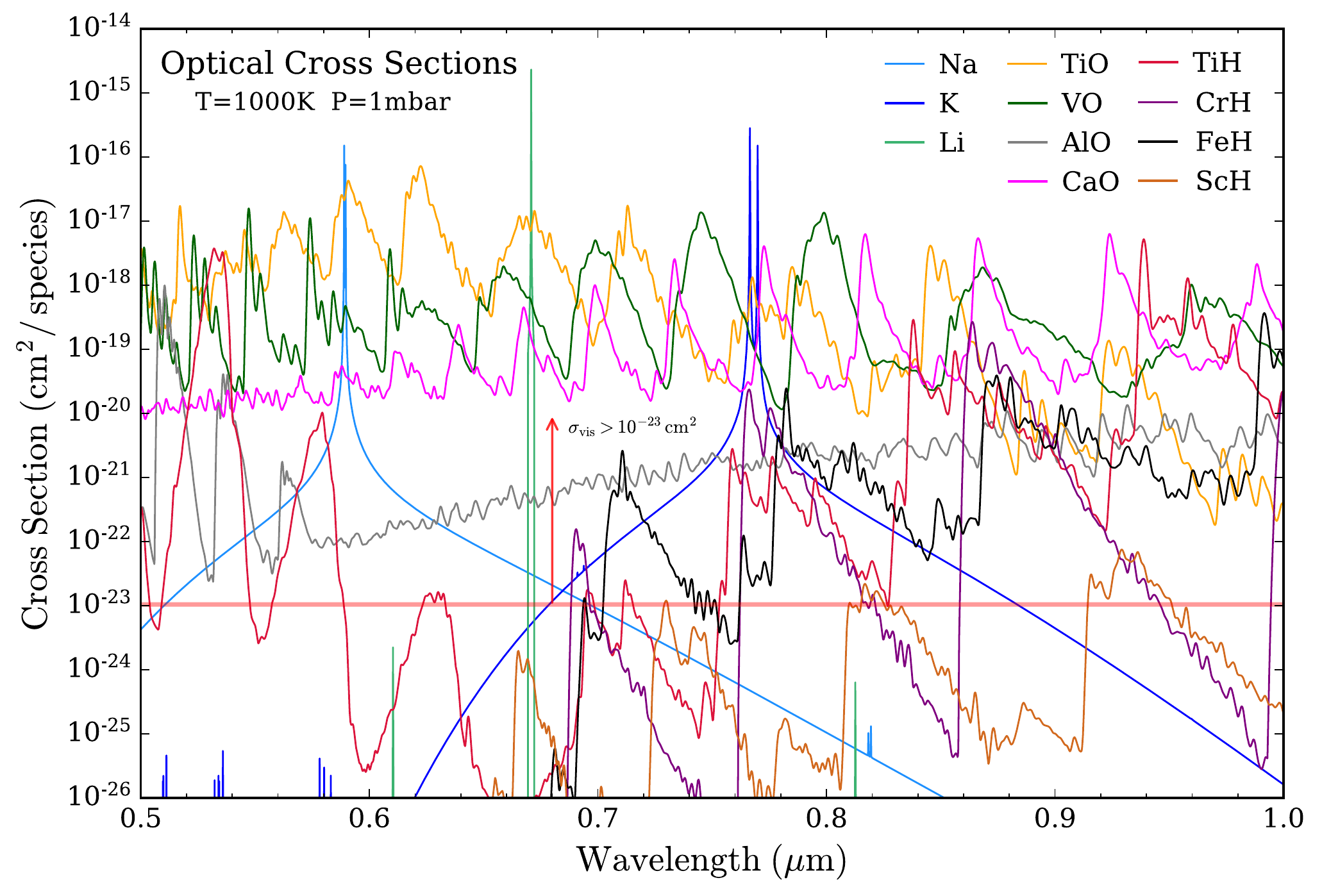} \caption{\textbf{Visible wavelength absorption cross sections}. The cross sections shown are calculated at 1000K and 1mbar, representative of the equilibrium temperature of HAT-P-26b and the photospheric pressure probed during transit. Metal oxide and hydride cross sections have been smoothed to $R \sim 1000$ for clarity, whilst alkali cross sections are plotted at $R=5000$ to clearly resolve resonance doublets. The criterion used to select prominent optical absorbers is indicated by the red shading and arrow.}
	\vspace{-1.0cm}
    \label{fig:cross_sections}
\end{figure}

We consider many prospective sources of infrared and visible opacity. Standard carbon, oxygen, and nitrogen-bearing species with well-known infrared absorption features are included: H$_2$O, CH$_4$, NH$_3$, HCN, CO, CO$_2$, and C$_2$H$_2$. H$_2$ and He are treated as a single gaseous species, with a fixed solar H$_2$/He ratio of 0.17 assumed. Given the suggestions of substructure in HAT-P-26b's visible-wavelength transmission spectrum (Figure \ref{fig:data}), we also consider chemical species with prominent optical cross sections. We therefore include alkali metals along with metal oxides and hydrides \citep{Sharp2007}. Specifically, we consider: Na, K, Li, TiO, VO, AlO, CaO, TiH, CrH, FeH, and ScH. Absorption cross sections for these species are shown in Figure \ref{fig:cross_sections}. The criterion for including an optical absorber in our models was taken to be $\sigma_{\mathrm{vis}} > 10^{-23}$ cm$^2$ at the equilibrium temperature of HAT-P-26b -- where `vis' indicates wavelengths covering the visible portion of the observations ($\sim 0.5-1.0 \, \micron$). This somewhat arbitrary criterion renders the dimensionality of the parameter space tractable for exploration by removing species with negligible absorption cross sections \citep[see][for a visual demonstration]{Tennyson2018}. We have verified that including additional species with weaker cross sections do not modify our results.

The molecular and atomic cross section database employed by POSEIDON has recently undergone a substantial upgrade. Compared to the cross sections used in \citet{MacDonald2017a}, the spectral resolution has been increased by a factor of 100 (to $\Delta \nu$ = 0.01 cm$^{-1}$), the temperature and pressure coverage has been extended ($10^{-6} - 10^{\, 2}$ bar and $100 - 3500$ K, respectively), and pressure broadening now accounts for H$_2$ and He (in an 85$\%$ to 15$\%$ mixture) instead of air broadening. The line list sources have also been revised to reflect the current state of the art. We no longer use line lists from the HITEMP-2010 database, opting instead for the more recent POKAZATEL H$_2$O line list \citep{Polyansky2018}, the \citet{Li2015} CO line list, and CDSD-4000 for CO$_2$ \citep{Tashkun2011}. We have also upgraded CH$_4$ to the latest 34to10 EXOMOL line list \citep{Yurchenko2017}, C$_2$H$_2$ to ASD-1000 \citep{Lyulin2017}, and the atomic transitions to VALD3 \citep{Pakhomov2017}. Line lists for the remaining molecules are taken from ExoMol \citep{Tennyson2016}, in particular \citet{Burrows2005,Burrows2002} for TiH and CrH and \citet{Lodi2015} for ScH. The calculation of these cross sections follows a similar method to \citet{Hedges2016} and \citet{Gandhi2017}.

We also include collisionally-induced absorption (CIA) and Rayleigh scattering. CIA due to H$_2$-H$_2$, H$_2$-He, and H$_2$-CH$_4$ is taken from HITRAN \citep{Richard2012}, whilst Rayleigh scattering cross section for H$_2$, He, and H$_2$O are derived from refractive indices in various sources \citep{Hohm1993,Mansfield1969,Hill1986}.

\subsubsection{Treatment of clouds and hazes} \label{subsubsection:cloud_opac}

An important source of opacity can be provided by clouds and hazes. Here, we consider clouds to constitute an opaque deck located at $P_{\mathrm{cloud}}$, below which no electromagnetic radiation may pass. This effectively corresponds to the grey, large particle size limit of Mie scattering \citep[e.g][]{Kitzmann2018}. Hazes, however, are distributed uniformly throughout the atmosphere with an extinction coefficient given by a two-parameter power law \citep{DesEtangs2008}: $\kappa_{\mathrm{haze}} = n_{\mathrm{H_{2}}} \, a \, \sigma_{0} (\lambda/\lambda_{0})^{\gamma}$, where $\lambda_0$ is a reference wavelength (350 nm), $\sigma_0$ is the $\mathrm{H}_2$-Rayleigh scattering cross section at the reference wavelength ($5.31 \times 10^{-31} \, \mathrm{{m}^2}$), $a$ is the `Rayleigh enhancement factor', and $\gamma$ is the `scattering slope'. This `enhanced-Rayleigh' slope accounts for the effect of small particle sizes, with the parameter $\gamma$ in principle indicative of the aerosol causing the slope \citep{Pinhas2017}. We additionally allow for inhomogeneous terminator cloud and haze distributions \citep{Line2016b,MacDonald2017a}, with a terminator coverage fraction $\bar{\phi}$. The effect of this `patchy cloud' on the radiative transfer calculation is espoused in the next section.

\subsection{Radiative transfer} \label{subsection:radiative_transfer}

Transmission spectra are computed by solving the equation of radiative transfer under a plane-parallel geometry. Here we outline the essential aspects, with a comprehensive discussion given in \citet{MacDonald2017a}. The slant optical depth for a given impact parameter is obtained by integrating the atmospheric extinction coefficient -- including chemical and cloud / haze opacity -- along the observer's line of sight. The effective planetary radius of an axially-symmetric atmosphere, at a given wavelength, is computed by integrating the atmospheric transmission, $e^{-\tau_{\lambda}}$, over successive annuli:
\begin{equation}\label{eq:transit_depth_1D}
R_{\mathrm{eff}, \, \lambda}^{2} = R_{\mathrm{top}}^{2} - \displaystyle\int_{0}^{R_{\mathrm{top}}} e^{-\tau_{\lambda}(b)} \, 2 b \, db
\end{equation} 
where $R_{\mathrm{top}}$ is the radial extent of the modelled atmosphere, $\tau_{\lambda}$ is the slant optical depth at wavelength $\lambda$, and $b$ is the impact parameter of a given ray. A `one-dimensional' transmission spectrum can then be computed: $\delta_{\lambda} \equiv (R_{\mathrm{eff, \, \lambda}}/{R_{*}})^{2}$, where $R_{*}$ is the stellar radius ($0.788 \, R_{\sun}$ for HAT-P-26).

Azimuthal inhomogeneities are included via a linear superposition of transit depths, weighted by the coverage fraction of a given region. Here, we consider the possibility of inhomogeneous clouds \citep{Line2016b} and hazes by superimposing cloudy and cloud-free models:
\begin{equation}\label{eq:transit_depth_2D}
\Delta_{\lambda} = \bar{\phi} \, \delta_{\lambda, \mathrm{cloudy}} + (1 - \bar{\phi}) \, \delta_{\lambda, \mathrm{clear}}
\end{equation} 
where $\bar{\phi}$ is the terminator cloud coverage fraction. This two-dimensional cloud / haze prescription facilitates the breaking of many cloud-haze-composition degeneracies that can manifest in strictly one-dimensional cloud models \citep{Benneke2013,MacDonald2017a}.

We evaluate transmission spectra at a constant spectral resolution of $R = 2000$ from 0.4-5.2 $\mathrm{\mu}$m. Model spectra are convolved with the relevant grism point spread functions and integrated over instrument response curves to produce binned model points at the resolution of the observations. Spitzer model points are produced by integrating over the IRAC 3.6 $\mathrm{\mu}$m and 4.5 $\mathrm{\mu}$m photometric instrument functions.

\subsection{Atmospheric parameterisation} \label{subsection:parameters}

\newcommand{\ra}[1]{\renewcommand{\arraystretch}{#1}}
\begin{table}
\ra{1.3}
\caption{Retrieval parameters and prior probabilities}
\begin{tabular*}{\columnwidth}{l@{\extracolsep{\fill}} llll@{}}\toprule
$\mathrm{Parameter}$ & $\mathrm{Prior}$ & $\mathrm{Range}$\\ \midrule
\textbf{P-T profile} & \\ 
\hspace{0.5em} $\alpha_{1,2}$ & Uniform & $0.02$ -- $2.0 \ \mathrm{K}^{-1/2}$  \\
\hspace{0.5em} $P_{1,2}$ & Log-uniform & $10^{-7}$ -- $10^{\,2} \ \mathrm{bar}$ \\
\hspace{0.5em} $P_{3}$ & Log-uniform & $10^{-2}$ -- $10^{\,2} \ \mathrm{bar}$ \\
\hspace{0.5em} $P_{\mathrm{ref}}$ & Log-uniform & $10^{-7}$ -- $10^{\,2} \ \mathrm{bar}$ \\
\hspace{0.5em} $T_{1 \mathrm{mbar}}$ & Uniform & $100$ -- $1400 \ \mathrm{K}$ \\ \midrule
\textbf{Composition} & \\ 
\hspace{0.5em} $X_{i}$ & Log-uniform & $10^{-14}$ -- $10^{-0.3}$ \\ \midrule
\textbf{Clouds} & \\
\hspace{0.5em} $a$ & Log-uniform & $10^{-4}$ -- $10^{\,8}$ \\
\hspace{0.5em} $\gamma$ & Uniform & \hspace{-1.0em} $-20$ -- $2$ \\
\hspace{0.5em} $P_{\mathrm{cloud}}$ & Log-uniform & $10^{-6}$ -- $10^{\,2} \ \mathrm{bar}$ \\
\hspace{0.5em} $\bar{\phi}$ & Uniform & $0 - 1$ \\ \midrule
\textbf{Other} & \\
\hspace{0.5em} $\delta_{\mathrm{rel}}$ & Uniform & \hspace{-1.0em} $-1000$ -- $1000$ ppm \\
\bottomrule
\vspace{0.1pt}
\end{tabular*}
\label{table:priors}
\end{table}

\begin{figure*}
	\includegraphics[width=\linewidth,  trim={-0.6cm 0.6cm -0.6cm 0.4cm}]{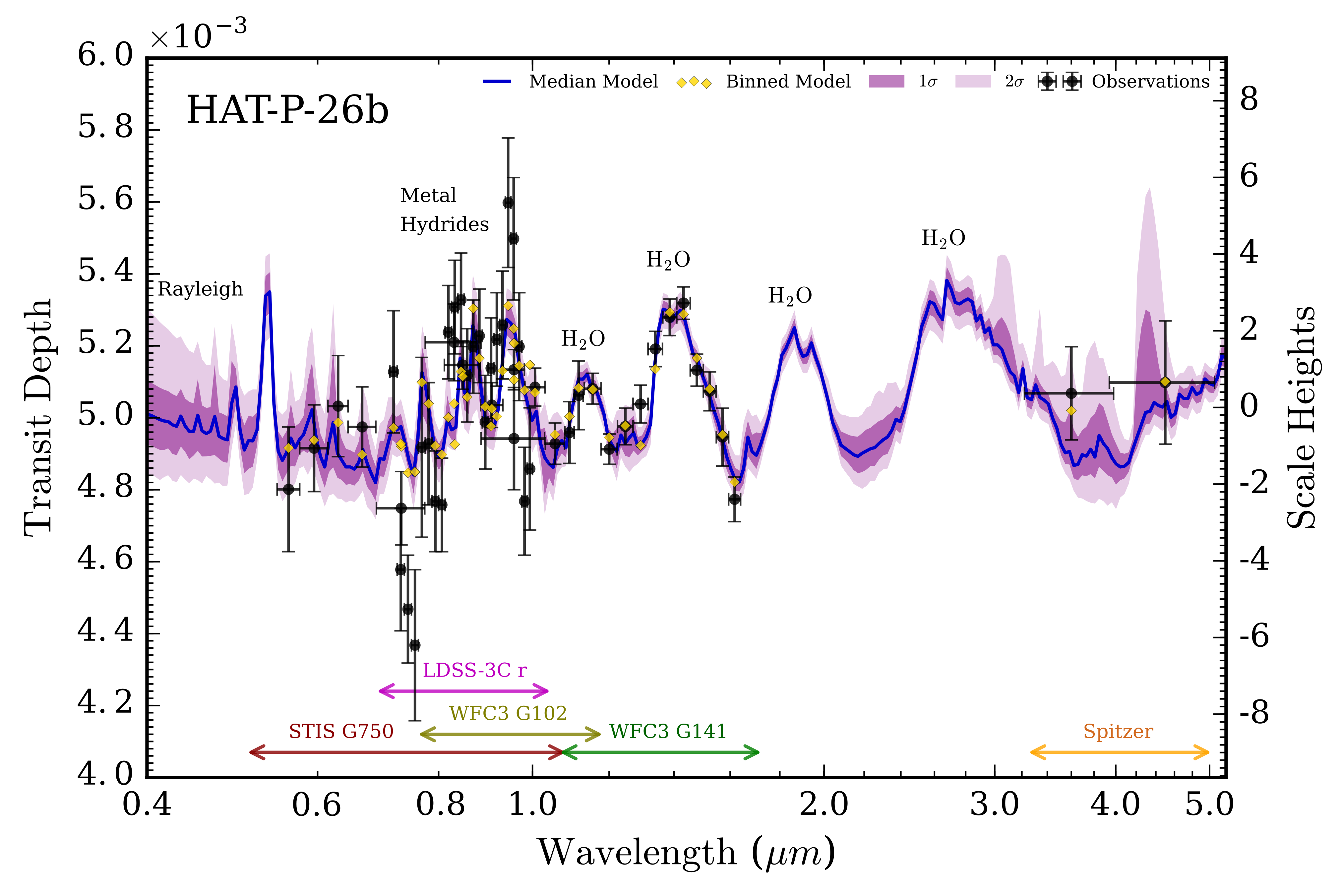}
    \caption{\textbf{Retrieved transmission spectrum of HAT-P-26b}. The observed transit depths are shown by black circles, with corresponding instrument modes and spectral ranges shown. A relative offset of +116ppm has been applied to the LDSS-3C red observations, based on the median retrieved $\delta_{\mathrm{rel}}$. The median model spectrum obtained by POSEIDON is shown in blue (at $R = 100$, for clarity), along with 1$\sigma$ and 2$\sigma$ confidence regions in purple, derived from 30,000 random posterior samples. The best-fitting model, binned to the resolution of the data, is shown by gold diamonds. Prominent absorption and scattering features are labelled.}
    \label{fig:spectrum_retrieved}
\end{figure*}

Each model atmosphere is expressed in terms of a parametric state vector, which encodes the pressure-temperature profile, atmospheric composition, and cloud / haze properties. The pressure-temperature (P-T) profile comprises either the flexible 6 parameter shape from \citet{Madhusudhan2009} (re-expressed in terms of the 1mbar temperature) or an isotherm, supplemented by $P_{\mathrm{ref}}$ -- the \emph{a priori} unknown pressure at $R_p$. The atmospheric composition (section \ref{subsubsection:chem_opac}) is encoded via up to 18 volume mixing ratios, $X_{i}$, assumed to be uniform both in altitude and across the terminator. Clouds and hazes are described by the 4 parameter prescription of \citet{MacDonald2017a} (section \ref{subsubsection:cloud_opac}). A relative offset between the \citet{Stevenson2016} and \citet{Wakeford2017} observations, $\delta_{\mathrm{rel}}$, is prescribed, as discussed in section \ref{section:observations}. Our models thus have a maximum of 30 free parameters, summarised in Table \ref{table:priors}.

Priors for each parameter are taken as either uniform or uniform-in-the-logarithm, depending on whether the range is less than or more than two orders of magnitude, respectively. The P-T profile and cloud parameter priors are chosen to be generous and uninformative, for the reasoning discussed in \cite{MacDonald2017a}. Mixing ratios have an upper limit of $10^{-0.3}$ ($\approx$ 50\%), with the remainder assumed to comprise H$_2$ and He in solar proportions. Strictly speaking, truly permutation invariant mixing ratios (e.g. equal a priori probabilities for a CO$_2$ or H$_2$O dominated atmosphere vs. a H$_2$-He dominated atmosphere) requires usage of a flat Dirchlet prior (uniform over the unit simplex) \citep{Benneke2012}. This approach is, however, computationally expensive for $\gtrsim 8$ mixing ratios, so we implicitly assume that the atmosphere of HAT-P-26b does not have $X_{\mathrm{H_{2}+He}} < 10^{-0.3}$ via our usage of bounded log-uniform priors on the other species. We have verified that usage of a Dirchlet prior results in no notable differences in the retrieved parameters for this planet.

\subsection{Retrieval methodology} \label{subsection:statistics}

POSEIDON is capable of both Bayesian parameter estimation and model comparison. For reporting parameter constraints, we utilise the complete model including all 30 parameters described in section \ref{subsection:parameters} -- this ensures any degeneracies, such as between clouds and composition, are captured via posterior marginalisation. As a guiding principle, we also employ nested Bayesian model comparison using the computed Bayesian evidences of competing models to identify the simplest model compatible with the observations \citep[see][for a full description of statistical aspects of our retrieval methodology]{MacDonald2017a}. This enables direct assessment of model complexity via evaluating the statistical significance of each model component (e.g. detection significances of chemical species, clouds, etc).

We conducted over 50 atmospheric retrievals in the course of our investigation. Each multi-dimensional parameter space is explored via the multimodal nested sampling algorithm MultiNest \citep{Feroz2008,Feroz2009}, as implemented by the python wrapper PyMultiNest \citep{Buchner2014}. Our retrievals typically employ 4000 MultiNest live points, with a Bayesian evidence termination error of $\Delta \ln \mathcal{Z} \approx 0.05$. A typical 30 dimensional retrieval then results in the computation of $\sim 5 \times 10^{6}$ transmission spectra. Using this methodology, we now proceed to describe constraints on the atmospheric properties of HAT-P-26b, including its metallicity, before assessing the degree of model complexity current observations warrant. 

\section{The Atmosphere of HAT-P-26b}\label{section:results}

We here present our inferences into the atmospheric properties of HAT-P-26b. We begin in section \ref{subsection:results_spectrum} with a brief discussion on the overall quality of our spectral fits to the observations. We then examine the atmospheric composition of HAT-P-26b in section \ref{subsection:results_composition}, constraints on clouds and the pressure-temperature (P-T) profile in sections \ref{subsection:results_clouds} and \ref{subsection:results_PT}, respectively, before examining the minimal model complexity necessary to adequately fit the observations in section \ref{subsection:results_model_complexity}. We detail in appendix \ref{section:Without_ground_data} an equivalent retrieval analysis utilising only observations from \citet{Wakeford2017}.

\subsection{Retrieved transmission spectrum} \label{subsection:results_spectrum}

Figure~\ref{fig:spectrum_retrieved} shows the retrieved transmission spectrum of HAT-P-26b. This corresponds to the `full' 30-dimensional model described in section \ref{subsection:parameters}, including a wide range of chemical species, non-isothermal P-T profiles, clouds and hazes, and a relative offset between the \citet{Stevenson2016} and \citet{Wakeford2017} datasets. The posterior distribution from this retrieval is available as supplementary online material. Our best-fitting model spectrum, binned to the resolution of the observations, lies within 1$\sigma$ for 66\% of the observations (33 out of 50, yielding $\chi_{r}^{2} = 3.55$). Much of the spectral structure in the infrared is explained by H$_2$O, as concluded by \citet{Wakeford2017}. The attribution of spectral features at visible wavelengths, well-fit by our models, to specific species is detailed in section \ref{subsubsection:detections}.

Before deriving atmospheric properties, we first assessed if a relative offset, $\delta_{\mathrm{rel}}$, was necessary. We ran an identical retrieval with the raw data alone, and conducted a Bayesian model comparison between this model and the full model. The result was a Bayes factor of 7.86 in favour of the full model (equivalent to 2.6$\sigma$), hence we include $\delta_{\mathrm{rel}}$ as a parameter in all subsequent retrievals. The likely reason for the retrieved offset, $\delta_{\mathrm{rel}} = 116 \pm 35$ ppm, is the likelihood penalty induced by the cluster of LDSS-3C observations around 0.72 $\micron$ (and to a lesser extent around 0.99 $\micron$) that are much lower than the other data points and are difficult to explain with our models. For example, the data around 0.72 $\micron$, are far lower than would be expected of pure H$_2$ Rayleigh scattering without any additional sources of opacity. \citet{Stevenson2016} has previously identified these points as outliers, which can potentially be attributed to fringing \citep{Wakeford2017}. Nevertheless, the retrieved parameters are consistent even when not using a relative offset. We turn now to atmospheric inferences from our spectral retrievals.

\subsection{Composition} \label{subsection:results_composition}

We first describe the atmospheric composition of HAT-P-26b obtained from our retrieval analysis. We establish in section \ref{subsubsection:detections} which chemical species included in our models are indicated by the observations, along with their associated statistical significances. We examine spectral features of each indicated species in section \ref{subsubsection:signatures}. Constraints on the abundances of these species are reported in section \ref{subsubsection:abundances}. Finally, in section \ref{subsubsection:derived}, constraints on derived atmospheric properties, such as the metallicity and C/O are presented.

\subsubsection{Detections} \label{subsubsection:detections}

We establish at high significance that the atmosphere of HAT-P-26b is primarily H$_2$+He dominated, with H$_2$O as a secondary component. We detect H$_2$O at $7.2\sigma$ confidence via a nested Bayesian model comparison (see below), in agreement with the previous detection of H$_2$O by \citet{Wakeford2017}. Notably, H$_2$O is the only considered species that explains the broad $\sim 1.2 \, \micron$ and $\sim 1.4 \, \micron$ absorption features (see Figure \ref{fig:spectrum_retrieved}). Following the confirmation of H$_2$O, we directly computed the significance of H$_2$+He as the background gas by considering a model with no H$_2$+He, H$_2$O as the dominant gas, and all other species as trace gases. This model poorly fit the observations -- primarily due to the resulting high mean molecular weight of $\sim 18$ atomic mass units -- enabling us to establish at $7.6\sigma$ confidence that H$_2$+He must be the dominant background gas.

\begin{table}
\ra{1.3}
\caption[]{Bayesian model comparison: composition of HAT-P-26b}
\begin{tabular*}{\columnwidth}{l@{\extracolsep{\fill}} cccccl@{}}\toprule
$\mathrm{Model}$ & \multicolumn{1}{p{1cm}}{\centering \hspace{-0.4cm} Evidence \\ \centering $ \hspace{-0.2cm} \mathrm{ln}\left(\mathcal{Z}_{i}\right)$}  & \multicolumn{1}{p{1cm}}{\centering Best-fit \\ \centering $ \chi_{r, \mathrm{min}}^{2}$} & \multicolumn{1}{p{1.7cm}}{\centering \hspace{-0.3cm} Bayes \\ \hspace{-0.2cm} Factor \\ \centering $ \hspace{-0.2cm} \mathcal{B}_{0i}$}& \multicolumn{1}{p{1cm}}{\centering \hspace{-0.6cm} Significance \\ \centering \hspace{-0.4cm} of Ref.}\\ \midrule
\textbf{Full Chem} & $ 352.26 $ & $ 3.55 $ & Ref. & Ref.\\
\hspace{0.2 em} No H$_2$+He & $ 325.59 $ & $ 6.32 $ & $ 3.82 \times 10^{11} $ & $7.6 \sigma$ \\
\hspace{0.2 em} No H$_2$O & $ 328.12 $ & $ 6.03 $ & $ 3.03 \times 10^{10} $ & $7.2 \sigma$ \\
\hspace{0.2 em} No CH$_4$ & $ 352.41 $ & $ 3.39 $ & $ 0.86 $ & N/A \\
\hspace{0.2 em} No NH$_3$ & $ 352.64 $ & $ 3.37 $ & $ 0.68 $ & N/A \\
\hspace{0.2 em} No HCN & $ 352.45 $ & $ 3.44 $ & $ 0.82 $ & N/A \\
\hspace{0.2 em} No CO & $ 352.34 $ & $ 3.39 $ & $ 0.92 $ & N/A \\
\hspace{0.2 em} No CO$_2$ & $ 352.24 $ & $ 3.40 $ & $ 1.01 $ & N/A \\
\hspace{0.2 em} No C$_2$H$_2$ & $ 352.39 $ & $ 3.41 $ & $ 0.88 $ & N/A \\
\hspace{0.2 em} No Na & $ 352.70 $ & $ 3.41 $ & $ 0.64 $ & N/A \\
\hspace{0.2 em} No K & $ 352.97 $ & $ 3.37 $ & $ 0.49 $ & N/A \\
\hspace{0.2 em} No Li & $ 352.43 $ & $ 3.41 $ & $ 0.84 $ & N/A \\
\hspace{0.2 em} No TiO & $ 352.47 $ & $ 3.47 $ & $ 0.81 $ & N/A \\
\hspace{0.2 em} No VO & $ 352.29 $ & $ 3.41 $ & $ 0.35 $ & N/A \\
\hspace{0.2 em} No AlO & $ 352.25 $ & $ 3.40 $ & $ 1.01 $ & N/A \\
\hspace{0.2 em} No CaO & $ 352.59 $ & $ 3.44 $ & $ 0.71 $ & N/A \\
\hspace{0.2 em} No TiH & $ 347.45 $ & $ 4.08 $ & $ 122 $ & $3.6 \sigma$ \\
\hspace{0.2 em} No CrH & $ 351.16 $ & $ 3.54 $ & $ 3.01 $ & $2.1 \sigma$ \\
\hspace{0.2 em} No FeH & $ 352.43 $ & $ 3.46 $ & $ 0.84 $ & N/A \\
\hspace{0.2 em} No ScH & $ 351.57 $ & $ 3.48 $ & $ 2.00 $ & $1.8 \sigma$ \\
\hspace{0.2 em} No ScH or AlO & $ 350.25 $ & $ 3.72 $ & $ 7.44 $ & $2.5 \sigma$ \\
\hspace{0.2 em} No M-Oxides & $ 354.08 $ & $ 3.04 $ & $ 0.16 $ & N/A \\
\hspace{0.2 em} No M-Hydrides & $ 345.66 $ & $ 3.79 $ & $ 732 $ & $4.1 \sigma$ \\
\bottomrule
\vspace{0.1pt}
\end{tabular*}
$\textbf{Notes}:$ The `Full Chem' reference model includes chemical opacity due to H$_2$, He, H$_2$O, CH$_4$, NH$_3$, HCN, CO, CO$_2$, C$_2$H$_2$, Na, K, Li, TiO, VO, AlO, CaO, TiH, CrH, FeH, and ScH. The `No M-Oxides' model has TiO, VO, AlO, and CaO removed; the `No M-Hydrides' model has TiH, CrH, FeH, and ScH removed. The number of degrees of freedom (d.o.f), given by $N_{\mathrm{data}} - N_{\mathrm{params}}$, is 20 for the reference model ($N_{\mathrm{data}}$ = 50). $\chi_{r, \, \mathrm{min}}^{2}$ is the minimum reduced chi-square ($\chi^2$/d.o.f). The significance indicates the degree of preference for the reference model, highlighted in bold, over each alternative model. N/A indicates no (or negative) evidence ($\mathcal{B}_{\mathrm{ij}} \lesssim 1$) supporting a given chemical species.
\label{table:composition_models}
\end{table}

\begin{figure*}
	\includegraphics[width=\linewidth,  trim={0.4cm 1.2cm -0.4cm 0.4cm}]{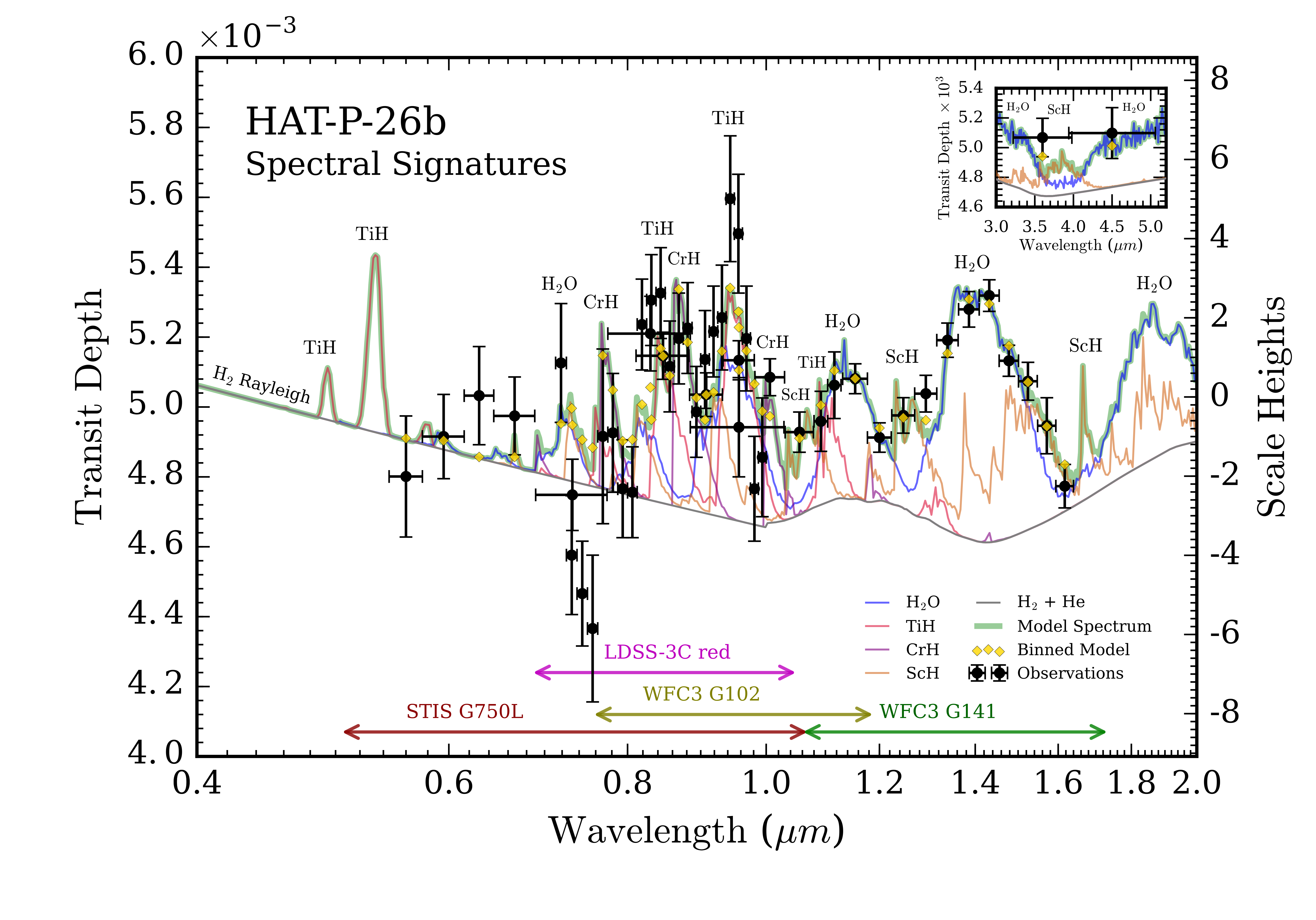}
	\caption{\textbf{Chemical signatures in the transmission spectrum of HAT-P-26b}. The green shading indicates the best-fitting spectrum, at $R=300$, from the `minimal complexity' model (see section \ref{subsection:results_model_complexity}). The grey line shows the spectral continuum from H$_2$ and He alone. Other lines display contributions of each retrieved chemical species above the H$_2$+He continuum. Prominent absorption and scattering features are labelled. The observed transit depths are shown by black circles, with corresponding instrument modes and spectral ranges shown. A relative offset of +125ppm has been applied to the LDSS-3C red observations, based on the median retrieved $\delta_{\mathrm{rel}}$. The best-fitting model, binned to the resolution of the data, is shown by gold diamonds. \textbf{Inset}: region surrounding Spitzer band-passes.}
    \label{fig:spectral_signatures}
\end{figure*}

The statistical significances of other chemical species were similarly established by nested Bayesian model comparisons. The results of this analysis are shown in Table \ref{table:composition_models}. The majority of potential trace gases have little or negative evidence supporting their presence ($\mathcal{B}_{\mathrm{ij}} \lesssim 1$). Bayes factors less than 1 indicate that a given chemical species is not necessary to explain the observations, as its inclusion results in `wasted' parameter space that penalises the Bayesian evidence \citep{Trotta2008}. There are, however, three exceptions: TiH ($\mathcal{B}_{\mathrm{ij}} = 122$ / 3.6$\sigma$), CrH ($\mathcal{B}_{\mathrm{ij}} = 3$ / 2.1$\sigma$), and ScH ($\mathcal{B}_{\mathrm{ij}} = 2$ / 1.8$\sigma$). On the Jeffreys' scale, where Bayes factors of 3, 12, and 150 are set for `weak', `moderate', and `strong' detections, respectively \citep{Trotta2008}, this corresponds to moderate evidence supporting TiH, weak evidence for CrH, and marginal evidence for ScH. Note that these significances automatically account for the possibility of overlapping spectral features, as the nested model comparisons fully map the parameter space without the species of interest to explore alternative explanations.

The positive evidence supporting three metal hydrides motivated additional tests. We firstly sought to establish whether the retrievals indicated metal hydrides uniquely, or a combination of metal hydrides and metal oxides. This was motivated by the presence of `L-shaped' degeneracies in the posterior between some metal oxides and hydrides (a prominent example is between ScH and AlO, see the supplementary online material). We therefore ran a retrieval with all four metal oxides (TiO, VO, AlO, and CaO) removed, and another with all four metal hydrides (TiH, CrH, FeH, and ScH) removed. By removing metal oxides, the Bayesian evidence improved ($\mathrm{ln}\left(\mathcal{Z}_{i}\right)$: 352.26 $\rightarrow$ 354.08) and the minimum reduced chi-square decreased ($ \chi_{r, \mathrm{min}}^{2}$: 3.55 $\rightarrow$ 3.04). However, by removing metal hydrides the Bayesian evidence suffered a substantial penalty ($\mathrm{ln}\left(\mathcal{Z}_{i}\right)$: 352.26 $\rightarrow$ 345.66) and the minimum reduced chi-square increased due to the worse fit at visible wavelengths ($ \chi_{r, \mathrm{min}}^{2}$: 3.55 $\rightarrow$ 3.79). Taken together, this indicates no evidence for metal oxides, but strong evidence ($\mathcal{B}_{\mathrm{ij}} = 722$ / 4.1$\sigma$) in favour of including metal hydrides.

In appendix \ref{section:Without_ground_data}, we repeat this Bayesian model comparison with the data of \citet{Stevenson2016} omitted to assess the robustness of these statistical significances. We find: H$_2$+He (6.7$\sigma$), H$_2$O (9.2$\sigma$), CrH (2.2$\sigma$), ScH (1.8$\sigma$), and metal hydrides (2.4$\sigma$). TiH is not supported without the LDSS-3C observations. Though the statistical significance of metal hydrides is reduced by only using the space-based data, the independent corroboration from the ground-based data supporting the same scenario strengthens the case for this atmospheric interpretation.

\newpage

\subsubsection{Spectral signatures} \label{subsubsection:signatures}

\begin{figure*}
	\includegraphics[width=0.98\linewidth,  trim={-0.5cm 0.2cm -0.5cm 0.4cm}]{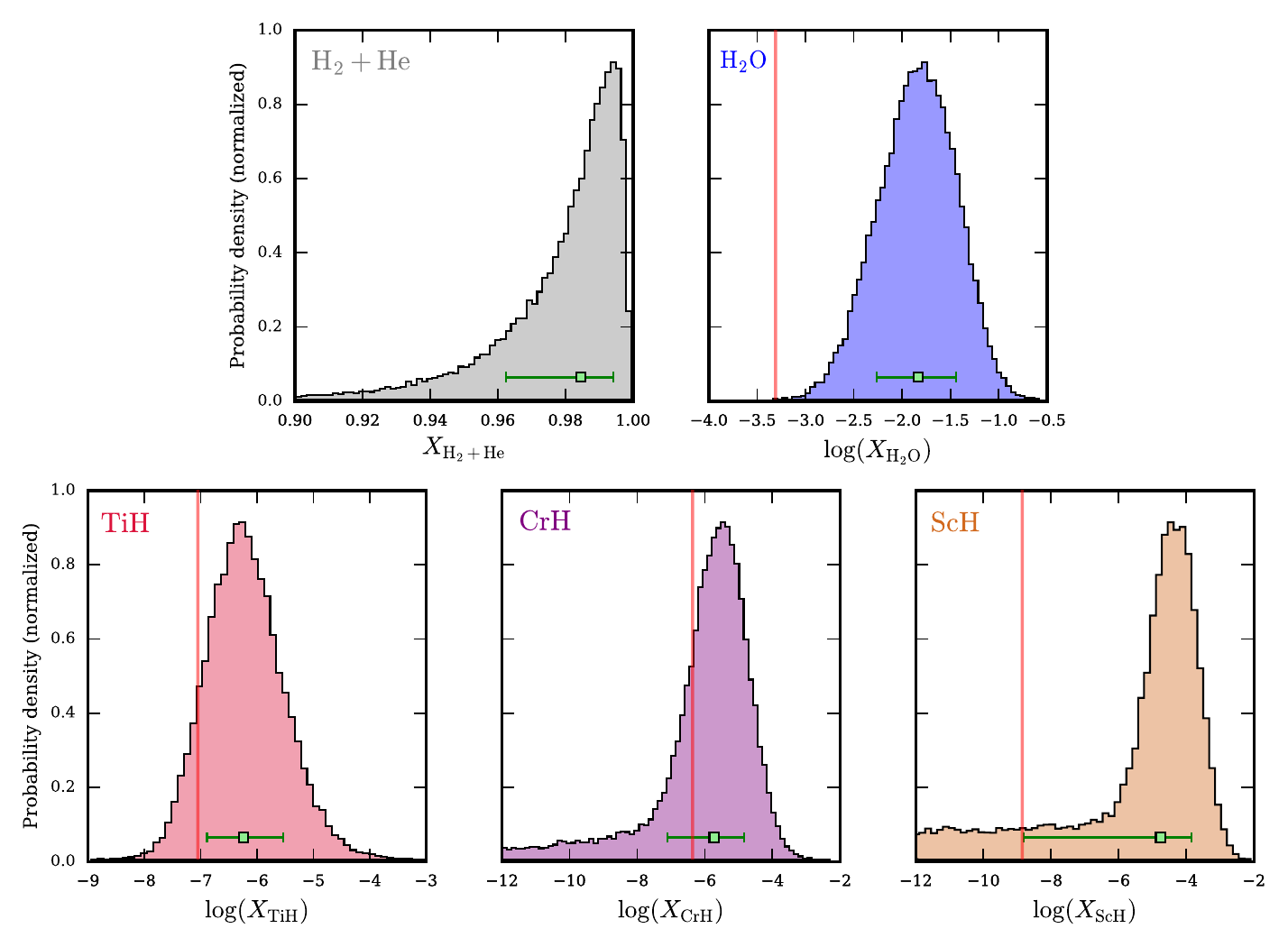}
	\caption{\textbf{Atmospheric composition of HAT-P-26b}. The histograms show the marginalised posterior probability distributions for the volume mixing ratios of each chemical species identified by the nested model comparison in Table \ref{table:composition_models}. H$_2$+He is detected at 7.6$\sigma$ and H$_2$O at 7.2$\sigma$. The trace gases TiH, CrH, and ScH are indicated at 3.6$\sigma$, 2.1$\sigma$, and 1.8$\sigma$, respectively. The combination of metal hydrides has a significance of 4.1$\sigma$. The abundance of the bulk component, H$_2$+He, is given in a linear scale, whilst the secondary component (H$_2$O) and trace gases are given on a log-scale. The median derived abundances and $\pm$1$\sigma$ confidence levels are denoted by green error bars. The vertical red lines indicate the solar O, Ti, Cr, and Sc abundances \citep{Asplund2009}.}
    \label{fig:abundances}
    \vspace{-12pt}
\end{figure*}

We investigate the attribution of specific observations to spectral signatures in Figure \ref{fig:spectral_signatures}. To highlight features of specific molecules, the spectrum shown only includes opacity from species with Bayes factors clearly exceeding unity in Table \ref{table:composition_models} (corresponding to the best-fitting spectrum from the `minimum complexity' model discussed in section \ref{subsection:results_model_complexity}). The spectrum is underlain by a continuum provided by H$_2$ Rayleigh scattering in the visible and H$_2$+He collisionally-induced absorption in the infrared. Above this continuum, the contributions of specific molecules are shown, with especially prominent features labelled.

The elevated infrared transit depths are best explained by the combination of H$_2$O and ScH. Specifically, the broad absorption features around 1.15 and 1.4$\micron$ in the Hubble WFC3 G141 band-pass are attributed to H$_2$O, whilst elevated absorption in windows between H$_2$O features (1.06, 1.25, and 1.6$\micron$) are attributed to opacity from ScH. There is also a minor contribution from TiH around 1.09$\micron$. A similar picture holds in the Spitzer band-passes, where H$_2$O features occur around 3.0 and 4.5$\micron$, with a smaller contribution from ScH around 3.9$\micron$. In particular, the low transit depth of the 4.5$\micron$ Spitzer IRAC point is the reason that CO or CO$_2$ are not detected. Despite these contributions to the overall spectral fit, the unique statistical significance of ScH is low (1.8$\sigma$). This is partly due to a degeneracy with AlO, which has many overlapping spectral features in the near-infrared with ScH (not shown), providing an alternative explanation for excess absorption between H$_2$O features. We verified this by running a retrieval with neither ScH nor AlO present, finding an increased significance of 2.5$\sigma$ for the combination. Note that these significances include full marginalisation over the possibility of clouds alternatively contributing this opacity (discussed further in section \ref{subsection:results_clouds}).

The visible wavelength transit depths yield the most compelling evidence for metal hydrides. In particular, 15 data points from $\sim$0.84-1.0$\micron$ are elevated by at least 1$\sigma$, and in some cases 3$\sigma$, above what can be explained by H$_2$+He and H$_2$O opacity alone (blue line, Figure \ref{fig:spectral_signatures}). We note that the independent LDSS-3C data \citep{Stevenson2016} and Hubble WFC3 G102/G141 data \citep{Wakeford2017} find good concordance over this spectral region, which contributes to the $>$4$\sigma$ combined evidence for the presence of metal hydrides. Regarding specific prominent features, TiH arises from the elevated absorption around 0.84 and 0.96$\micron$, whilst CrH contributes around 0.78, 0.88, and 1.02$\micron$. The inference of TiH is primarily caused by the elevated LDSS-3C points. The best-fitting spectrum offers a testable prediction for the presence of TiH: a search for elevated absorption from the strong TiH feature at 0.54$\micron$. This could be accomplished with the Hubble STIS G430 grism.

\subsubsection{Abundance constraints} \label{subsubsection:abundances}

Constraints on the abundances of each constituent of HAT-P-26b's atmosphere are presented in Figure \ref{fig:abundances}. We show only the posterior distributions for species whose presence was supported by the Bayesian model comparison in section \ref{subsubsection:detections}. These histograms are the result of marginalising over a wide range of potential atmospheric scenarios, including: degeneracies due to overlapping spectral features, clouds, reference-pressure normalisation, and non-isothermal temperature structures. This ensures that the quoted constraints are as conservative as our model assumptions allow. For non-detected species, upper bounds may still be placed on their abundances through ruling out values above which spectral features not apparent in the observations would have manifested -- these 2$\sigma$ upper bounds are given in a table inset in the supplementary online material.

The derived bounded abundance constraints are as follows. The most well-constrained mixing ratios are those of H$_2$+He and H$_2$O, correspondingly: $X_{\mathrm{H_2+He}} = 0.985^{+0.010}_{-0.022}$ and $\mathrm{log}(X_{\mathrm{H_2O}}) = -1.83^{+0.39}_{-0.43}$. The latter, with a median precision of 0.41 dex, represents the most precise water abundance measurement for an exo-Neptune to date. For the inferred metal hydrides, the mixing ratio constraints are: $\mathrm{log}(X_{\mathrm{TiH}}) = -6.24^{+0.71}_{-0.65}$, $\mathrm{log}(X_{\mathrm{CrH}}) = -5.72^{+0.89}_{-1.37}$, and $\mathrm{log}(X_{\mathrm{ScH}}) = -4.76^{+0.91}_{-4.09}$. We note that the tail to lower abundances for CrH is caused by the data points around 0.8$\micron$ having sufficiently broad 1$\sigma$ uncertainties to encompass both the H$_2$+He continuum and the possible contribution of CrH (see Figure \ref{fig:spectral_signatures}). Similarly, the long tail in the ScH abundance posterior manifests due to a degeneracy with AlO absorption signatures, as discussed in section \ref{subsubsection:signatures}. These tails are reflected accordingly in the low marginalised significance for these two species (2.1$\sigma$ and 1.8$\sigma$, respectively).

The present-day solar photoshperic O, Ti, Cr, and Sc abundances from \citet{Asplund2009} ($\mathrm{log(M/H)}$ = -3.31, -7.05, -6.36, -8.85, respectively) are overlaid in Figure \ref{fig:abundances}. We see that the H$_2$O abundance is conclusively super-solar (3$\sigma$ lower limit of $\mathrm{log}(X_{\mathrm{H_2O}}) = -3.00$). The median abundance of TiH is enhanced by a factor of 6.5 over the solar Ti abundance (consistent to 2$\sigma$). CrH is consistent with the solar Cr abundance. The mode of the ScH posterior is enhanced by $\sim 10^{4}$ over the solar Sc abundance, though the tail remains consistent with the solar Sc abundance within 1$\sigma$. However, due to the low significance of ScH, we advise caution in inferences derived from this value. The plausibility of these inferred metal hydrides is discussed in detail in section \ref{subsection:metal_hydride_plausibility}. 

We briefly note that there has been considerable discussion in the literature regarding the ability to obtain precise absolute molecular abundance constraints from transmission spectra \citep[e.g.][]{Benneke2012,Griffith2014,Heng2017}. Concerns have been raised that retrievals of low-resolution near-infrared transmission spectra (particularly WFC3 data) can suffer from a degeneracy between a reference radius, the corresponding reference pressure, and molecular mixing ratios (i.e. a three-way degeneracy between $R_p$, $P_{\mathrm{ref}}$, and $X_{i}$). In our present study, however, we have a long spectral baseline from the optical through the infrared. This allows simultaneous constraints on $X_{i}$ and $P_{\mathrm{ref}}$ via measurements of molecular absorption features and an independent continuum. In particular, H$_2$ Rayleigh scattering in the optical and H$_2$-H$_2$ CIA in the near-infrared provide independent continua in their respective spectral ranges \citep{deWit2013,Line2016b}, as shown in Figure \ref{fig:spectral_signatures}, facilitating the breaking of this degeneracy. The corresponding posteriors (e.g. our Figure \ref{fig:posterior_isotherm} and \citet{Pinhas2019}, Fig. 7) hence show only a weak correlation between each $X_{i}$ and $P_{\mathrm{ref}}$. We thereby see that this degeneracy does not impede our ability to obtain high-precision abundance measurements of HAT-P-26b's atmosphere.

\vspace{0.2cm}

\subsubsection{Metallicity, mean molecular weight, and C/O} \label{subsubsection:derived}

\begin{figure*}
	\includegraphics[width=\linewidth,  trim={0.0cm 0.4cm 0.0cm 0.0cm}]{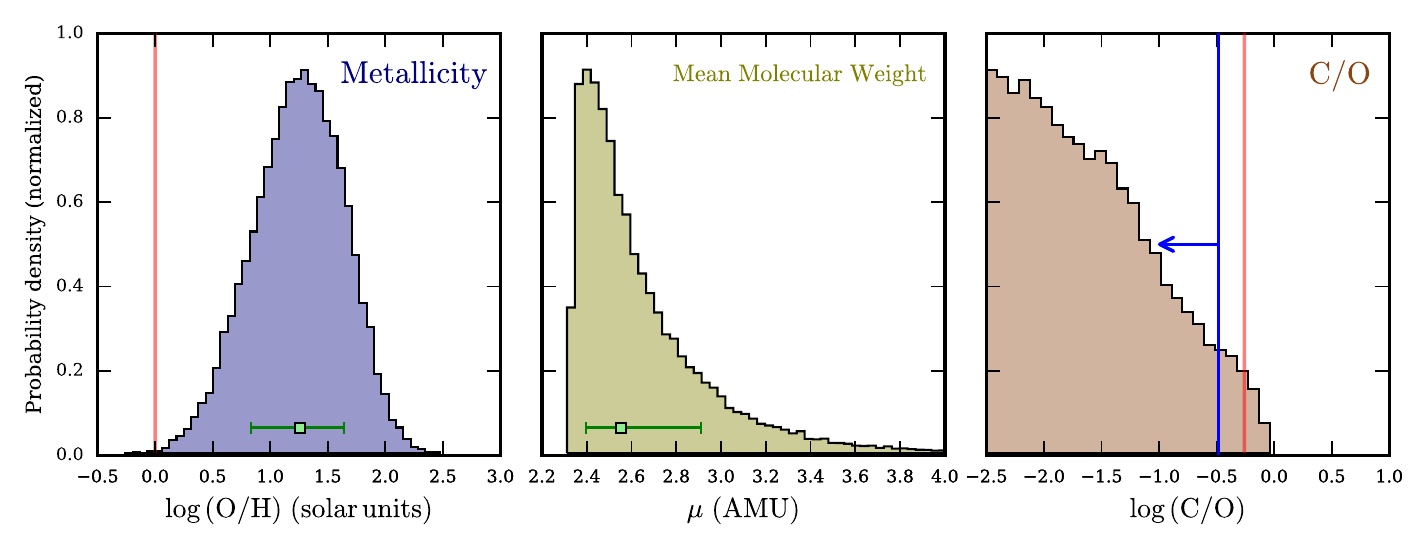}
	\caption{\textbf{Derived properties of HAT-P-26b's atmosphere}. The histograms show posterior distributions for the metallicity, mean molecular weight, and C/O ratio. Metallicity and C/O are given on a logarithmic scale. `AMU' denotes atomic mass units. Median derived values and $\pm$1$\sigma$ confidence levels are denoted by green error bars. The vertical red lines indicate solar metallicity and C/O.}
    \label{fig:derived}
\end{figure*}

We now present constraints on derived bulk properties of HAT-P-26b's atmosphere. From the posterior distribution of each chemical species, statistical estimates of the atmospheric metallicity, carbon-to-oxygen ratio, and mean molecular weight can be obtained. Metallicity is commonly defined as the number ratio of elements with atomic number $>$ 2 compared to that of hydrogen, relative to the solar value for the same elemental ratio (i.e. $\mathrm{\frac{M/H}{M/H_{\odot}}}$, where `M' is any element heavier than He). The mean molecular weight is the average mass of the chemical species comprising the atmosphere, commonly expressed in atomic mass units. The C/O elemental ratio provides an important diagnostic of the overall atmospheric composition \citep[e.g][]{Madhusudhan2012}.

\newpage

The only elemental ratio that has been robustly determined for all four of the solar system giant planets is C/H \citep[via the CH$_4$ abundance, see][]{Atreya2016}. This difficulty is due to other refractory species condensing out of the gas phase at low temperatures. C is thus taken as a proxy for `M' in deriving solar system giant metallicities. For hot Jupiters, whose higher temperatures yield spectra often dominated by H$_2$O opacity \citep{Sing2016}, the metallicity has instead been inferred via the O/H ratio of the atmosphere. In what follows, and throughout this paper, we similarly take the term `metallicity' to mean $\mathrm{\frac{O/H}{O/H_{\odot}}}$.

Previous studies have mainly taken two approaches to estimate atmospheric metallicities. Some studies treat the metallicity as a free, directly retrieved, parameter and, assuming chemical equilibrium, derive `chemically-consistent' mixing ratios \citep[e.g.][]{Fraine2014,Wakeford2017}. Others retrieve the H$_2$O mixing ratio directly from a spectrum and divide by the H$_2$O mixing ratio expected of a solar-composition atmosphere at the retrieved temperature under chemical equilibrium \citep[e.g.][]{Kreidberg2014}. The latter approach, often referred to as `free retrieval' does not directly impose chemical equilibrium in deriving the atmospheric H$_2$O mixing ratio itself, but implicitly assumes chemical equilibrium in metallicity estimation by dividing by a `solar' mixing ratio which does assume chemical equilibrium.

Here we take a different approach, free from the assumption of chemical equilibrium. We derive the O/H posterior probability distribution by drawing samples directly from the posterior distribution of each molecular mixing ratio. For a given sample:

\begin{equation}
\mathrm{O/H} = \frac{X_{\mathrm{H_{2}O}} + X_{\mathrm{CO}} + 2X_{\mathrm{CO_{2}}} + X_{\mathrm{TiO}} + ...}{2X_{\mathrm{H_{2}}} + 2X_{\mathrm{H_{2}O}} + 4X_{\mathrm{CH_{4}}} + 3X_{\mathrm{NH_{3}}} + X_{\mathrm{HCN}} + ...}
\end{equation} \label{eq:O_to_H}
where `...' on the numerator indicates all other species containing O atoms, and `...' on the numerator indicates all other H-bearing species. The metallicity is then obtained by dividing this elemental ratio by the present-day solar photosphere O/H ratio ($\mathrm{O/H}_{\odot} = 4.90 \times 10^{-4}$, \citet{Asplund2009}). When running a retrieval where one or more of these species is not included, their mixing ratios are set to 0. A posterior distribution for the metallicity is then constructed from the ensemble of O/H ratios derived from the samples. The C/O ratio is similarly constructed:

\begin{equation}\label{eq:C_to_O}
\mathrm{C/O} = \frac{X_{\mathrm{CH_{4}}} + X_{\mathrm{HCN}} + X_{\mathrm{CO}} + X_{\mathrm{CO_{2}}}}{X_{\mathrm{H_{2}O}} + X_{\mathrm{CO}} + 2X_{\mathrm{CO_{2}}} + X_{\mathrm{TiO}} + X_{\mathrm{VO}} + X_{\mathrm{AlO}}}
\end{equation}
though we note that present observations only place upper limits on HAT-P-26b's C/O ratio, due to the non-detection of carbon-bearing species. Finally, the atmospheric mean molecular weight is given by:

\begin{equation}\label{eq:mmw}
\mu = \sum_{i} \frac{X_{i} \, m_{i}}{m_{u}}
\end{equation}
where $m_{i}$ is the mass of each chemical species and $m_{u}$ is the atomic mass unit. For reference, hydrogen-dominated atmospheres with a solar H$_2$/He ratio have $\mu \approx 2.3$.

\newpage

The posterior distributions constructed in this manner are shown in Figure \ref{fig:derived}. These derived quantities from our full posterior are also provided in a table inset in the supplementary online material. An important caveat merits mention: only gas phase chemistry in the observable atmosphere is reflected by these derived properties; the formation of condensates at altitudes below those probed by the transmission spectrum could lead to the inferred values differing from the well-mixed values in the deep atmosphere.

We find the atmosphere of HAT-P-26b to be metal-enriched. Our derived metallicity, $18.1^{+25.9}_{-11.3} \times$ solar, revises the previous inference obtained from space-based spectra alone ($= 4.8^{+21.5}_{-4.0} \times$ solar, \citet{Wakeford2017}). Though these metallicities are consistent to 1$\sigma$, the higher median and increased precision of our value alters the interpretation. Specifically, whilst the previously inferred metallicity was consistent with a solar metallicity, we rule out this possibility to $>$3$\sigma$ confidence (lower limit, $1.3 \times$ solar). Our revised metallicity, the most precise obtained for an exo-Neptune to date, is lower than the solar system mass-metallicity expectation ($\sim 60 \times$ solar), but consistent to 2$\sigma$. We discuss the implications of these findings in light of the mass-metallicity diagram of other planets in section \ref{subsection:mass-metallicity}, and for the formation of HAT-P-26b in section \ref{subsection:formation_conditions}.

Constraints on the mean molecular weight and C/O from current observations are less informative. We find an atmosphere marginally heavier than a solar H$_2$-He mixture ($\mu = 2.55^{+0.36}_{-0.16}$), but consistent to within 2$\sigma$. The enhanced value is due to the molecular weight contribution of the $1.5^{+2.1}_{-0.9}\%$ H$_2$O comprising the atmosphere. Non-detections of carbon-bearing species place a 2$\sigma$ upper bound on the C/O ratio of 0.33, implying a sub-solar C/O (cf. 0.55). However, we caution that this conclusion is drawn primarily from the 4.5$\micron$ Spitzer point, and future spectroscopic infrared observations will be required to meaningfully constrain the C/O ratio (see section \ref{section:JWST}). Having presented constraints on HAT-P-26b's atmospheric composition, we now turn to inferences of the temperature structure and cloud properties.

\begin{figure*}
	\includegraphics[width=0.95\linewidth,  trim={-0.4cm 0.3cm -0.4cm 0.4cm}]{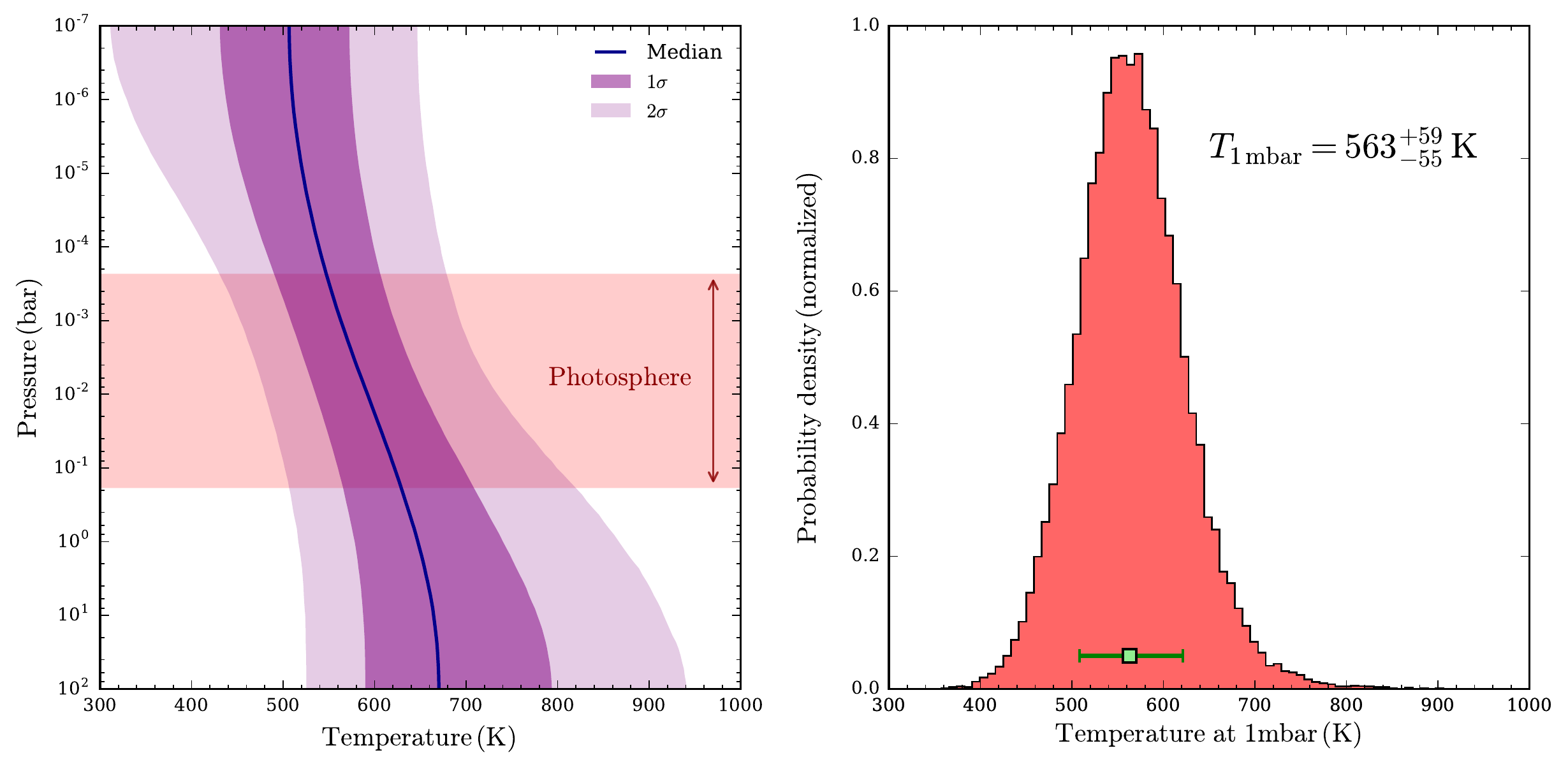}
	\caption{\textbf{Terminator temperature structure of HAT-P-26b}. Left: Retrieved pressure-temperature (P-T) profile. The median P-T profile obtained by POSEIDON from HAT-P-26b's transmission spectrum is shown in blue, with corresponding $1\sigma$ and $2\sigma$ confidence regions in purple, derived from 30,000 random posterior samples. The $\tau \approx 1$ infrared photosphere probed by the WFC3 observations is shaded in red. The pressures probed range from low pressures coinciding with maximal H$_2$O absorption (around the 1.432~$\micron$ datum), to the deep regions probed away from the H$_2$O absorption feature (around the 1.617~$\micron$ datum). Right: Marginalised posterior probability distribution of the temperature at the 1 mbar pressure level, serving as a proxy for the typical temperature probed in the line-of-sight.}
    \label{fig:PT}
\end{figure*}

\newpage

\subsection{Clouds and hazes} \label{subsection:results_clouds}

Our retrievals including clouds and hazes obtain relatively loose constraints on their underlying parameters. Specifically, we obtain a lower limit on the cloud top pressure ($\mathrm{log}(P_{\mathrm{cloud}}) > -4.28$) and an upper limit on the haze Rayleigh enhancement factor ($\mathrm{log}(a) < 4.72$) -- both to 2$\sigma$. The haze slope is essentially unconstrained. The cloud coverage fraction tends to prefer cloud-free models, with a long tail to higher fractions arising from the joint posterior in the $\mathrm{log}(P_{\mathrm{cloud}}) - \bar{\phi}$ plane following an L-shape bordering an exclusion zone (seen in the lower right corner of the posterior in the supplementary online material). The exclusion zone is caused by high-altitude clouds with low cloud coverage and low-altitude clouds with high cloud coverage both producing similar spectra. If the cloud coverage fraction is forced to be uniform, the loss of this degree of freedom yields a more stringent constraint on the cloud top pressure ($\mathrm{log}(P_{\mathrm{cld}}) \gtrsim -2.0$, not shown).

The retrieved cloud and haze properties can be understood as follows. The loose haze parameter constraints arise from the lack of a short wavelength slope in the optical data. Similarly, the substructure in the optical (for both the Hubble and Magellan datasets) is poorly fit by a constant transit depth, ruling out high-altitude uniform clouds. Models with high-altitude patchy clouds, which tend to broaden the wings of absorption features \citep[e.g.][]{MacDonald2017a}, are also not necessary to explain the widths of the near infrared H$_2$O features. We note that our non-inference of a cloud deck differs from the suggestion of clouds presented by \citet{Wakeford2017}, as they did not consider optical sources of molecular opacity in their models. We show in appendix \ref{section:Without_ground_data} that these conclusions also hold for retrievals without the ground-based observations. Taken together, these constraints suggest that opacity due to clouds or hazes is not necessary to explain the present observations.

\newpage

\subsection{Temperature structure} \label{subsection:results_PT}

Our retrieved P-T profile and 1 mbar temperature posterior are shown in Figure \ref{fig:PT}. We constrain $T_{\mathrm{1 mbar}}$ to $563^{+59}_{-55}$ K. This is cooler than the estimated equilibrium temperature of HAT-P-26b ($\approx$ 1000 K), which is consistent with the transmission spectrum probing the colder upper regions of the terminator. The $T_{\mathrm{1 mbar}}$ posterior (Figure \ref{fig:PT}, right) automatically accounts for variations in the local scale height ($k_B T / \mu g$) due to mean molecular weight variations, as the same draw of random samples are used to construct Figure \ref{fig:derived}. The reported constraints on $T_{\mathrm{1 mbar}}$ and $\mu$ therefore consistently account for one-another. We further note that the `scale height' axes shown on the right of our transmission spectra plots (e.g. Figure \ref{fig:spectrum_retrieved}) are expressed in terms of the median retrieved values of $T_{\mathrm{1 mbar}}$ and $\mu$.

The retrieved terminator P-T profile exhibits a temperature gradient across the photosphere. We illustrate the pressure range probed by the infrared observations ($\sim 10^{-4} - 10^{-1}$ bar, see Figure \ref{fig:PT}, left) by computing the $\tau \approx 1$ pressure probed by the centre of a prominent H$_2$O absorption feature and the neighbouring continuum -- specifically, pressures corresponding to the data points with the maximum (1.432 $\micron$) and minimum (1.617 $\micron$) infrared transit depths. Across the line-of-sight photosphere, the median P-T profile exhibits a temperature gradient of $\sim$ 80 K, though gradients as high as $\sim$ 200 K down to isothermal profiles are consistent with the retrieved 1$\sigma$ profile. Given this relatively shallow retrieved gradient, one might imagine that purely isothermal models are compatible with present observations. We now turn to rigorously asses this possibility, along with our prior suggestion that cloud and haze opacity appear absent, through a model complexity analysis.

\newpage

\subsection{Evaluating model complexity} \label{subsection:results_model_complexity}

We seek to establish the simplest atmospheric model capable of adequately explaining the transmission spectrum of HAT-P-26b. In the previous sections we presented constraints on the atmospheric composition, temperature structure, and cloud properties fully marginalised over a wide range of potential model complexity. Here, we shall asses the extent to which each aspect of this complexity is warranted. As discussed in section \ref{subsection:statistics}, our guiding principle here is nested Bayesian model comparison. This exercise can be viewed as identifying models which maximise the Bayesian evidence, or, equivalently, identifying those models with a reduced chi-square closer to 1. Given that our `full' retrieval had a reduced chi-square of 3.55 (Table \ref{table:composition_models}), it is clear that this 30 dimensional retrieval has many redundant free parameters, each of which we will now examine.

The optimal subset of chemical species is readily deduced. In section \ref{subsubsection:detections}, we found that only H$_2$+He, H$_2$O, TiH, CrH, and ScH have Bayes factors clearly in excess of 1. Thereby, we first conducted a retrieval containing only this reduced set of chemical species, with all other parameters (P-T, clouds, and offset) remaining. This reduced model, which we term `P-T + Clouds', has 16 free parameters, a Bayesian evidence of 356.06, and an improved $\chi_{r, \mathrm{min}}^{2}$ of 2.21 -- indicating that the 14 removed chemical species are redundant parameters. The remaining set of molecules constitutes the minimal set required to fit the observations.

We asses the role of temperature structure and clouds via Bayesian model comparisons. Two reference models are considered: (i) non-isothermal P-T profiles including clouds and hazes (`P-T + Clouds'); and (ii) isothermal atmospheres including clouds and hazes (`Iso + Clouds'). For each reference model, two additional retrievals were conducted: one without hazes, and one without clouds or hazes. These six retrievals range from 7-16 free parameters. The resulting model comparison is shown in Table \ref{table:PT_cloud_models}. We see that models without clouds and hazes have a higher or relatively unchanged Bayesian evidence, and a lower $\chi_{r, \mathrm{min}}^{2}$. This reinforces the finding of section \ref{subsection:results_clouds} that clouds and hazes minimally influence the retrieved spectra. The Bayesian evidence is maximised for the non-isothermal model with a clear atmosphere, though the preference over isothermal models is marginal ($\mathcal{B}_{\mathrm{ij}} = 1.1$ between the clear P-T model and the clear isotherm model). The lowest $\chi_{r, \mathrm{min}}^{2}$ is obtained for the isothermal clear model, for reasons we shall now explain.

\begin{table}
\ra{1.3}
\caption[]{Bayesian model comparison: importance of temperature structure and clouds in the atmosphere of HAT-P-26b}
\begin{tabular*}{\columnwidth}{l@{\extracolsep{\fill}} cccccl@{}}\toprule
$\mathrm{Model}$ & \multicolumn{1}{p{1cm}}{\centering \hspace{-0.4cm} Evidence \\ \centering $ \hspace{-0.2cm} \mathrm{ln}\left(\mathcal{Z}_{i}\right)$}  & \multicolumn{1}{p{1cm}}{\centering Best-fit \\ \centering $ \chi_{r, \mathrm{min}}^{2}$} & \multicolumn{1}{p{1.7cm}}{\centering \hspace{-0.3cm} Bayes \\ \hspace{-0.2cm} Factor \\ \centering $ \hspace{-0.2cm} \mathcal{B}_{0i}$}& \multicolumn{1}{p{1cm}}{\centering \hspace{-0.6cm} Significance \\ \centering \hspace{-0.4cm} of Ref.}\\ \midrule
\textbf{P-T + Clouds} & $ 356.06 $ & $ 2.21 $ & Ref. & Ref.\\
\hspace{0.2 em} No Haze & $ 356.30 $ & $ 2.10 $ & $ 0.79 $ & N/A \\
\hspace{0.2 em} Clear Skies & $ 356.32 $ & $ 1.97 $ & $ 0.77 $ & N/A \\
\midrule
\textbf{Iso + Clouds} & $ 356.24 $ & $ 1.94 $ & Ref. & Ref.\\
\hspace{0.2 em} No Haze & $ 356.23 $ & $ 1.84 $ & $ 1.00 $ & N/A \\
\hspace{0.2 em} Clear Skies & $ 356.22 $ & $ 1.75 $ & $ 1.02 $ & N/A \\
\bottomrule
\vspace{0.1pt}
\end{tabular*}
$\textbf{Notes}:$ Two reference models (bold) are considered: (i) `P-T' models including a variable pressure-temperature profile; (ii) `Iso' models with an isothermal temperature structure. The reference models include an optically thick cloud deck and a uniform-with-altitude haze (section \ref{subsubsection:cloud_opac}). All models include opacity due to H$_2$, He, H$_2$O, TiH, CrH, and ScH. The number of degrees of freedom (d.o.f), given by $N_{\mathrm{data}} - N_{\mathrm{params}}$, is 34 for the `P-T' reference model and 39 for the `Iso' reference model ($N_{\mathrm{data}}$ = 50). $\chi_{r, \, \mathrm{min}}^{2}$ is the minimum reduced chi-square ($\chi^2$/d.o.f). The significance indicates the degree of preference for the reference model, highlighted in bold, over each alternative model. N/A indicates no (or negative) evidence ($\mathcal{B}_{\mathrm{ij}} \lesssim 1$) supporting hazes or clouds.
\vspace{-0.2cm}
\label{table:PT_cloud_models}
\end{table}

\begin{figure}
	\includegraphics[width=\linewidth, trim={1.0cm 1.4cm 0.4cm 0.0cm}]{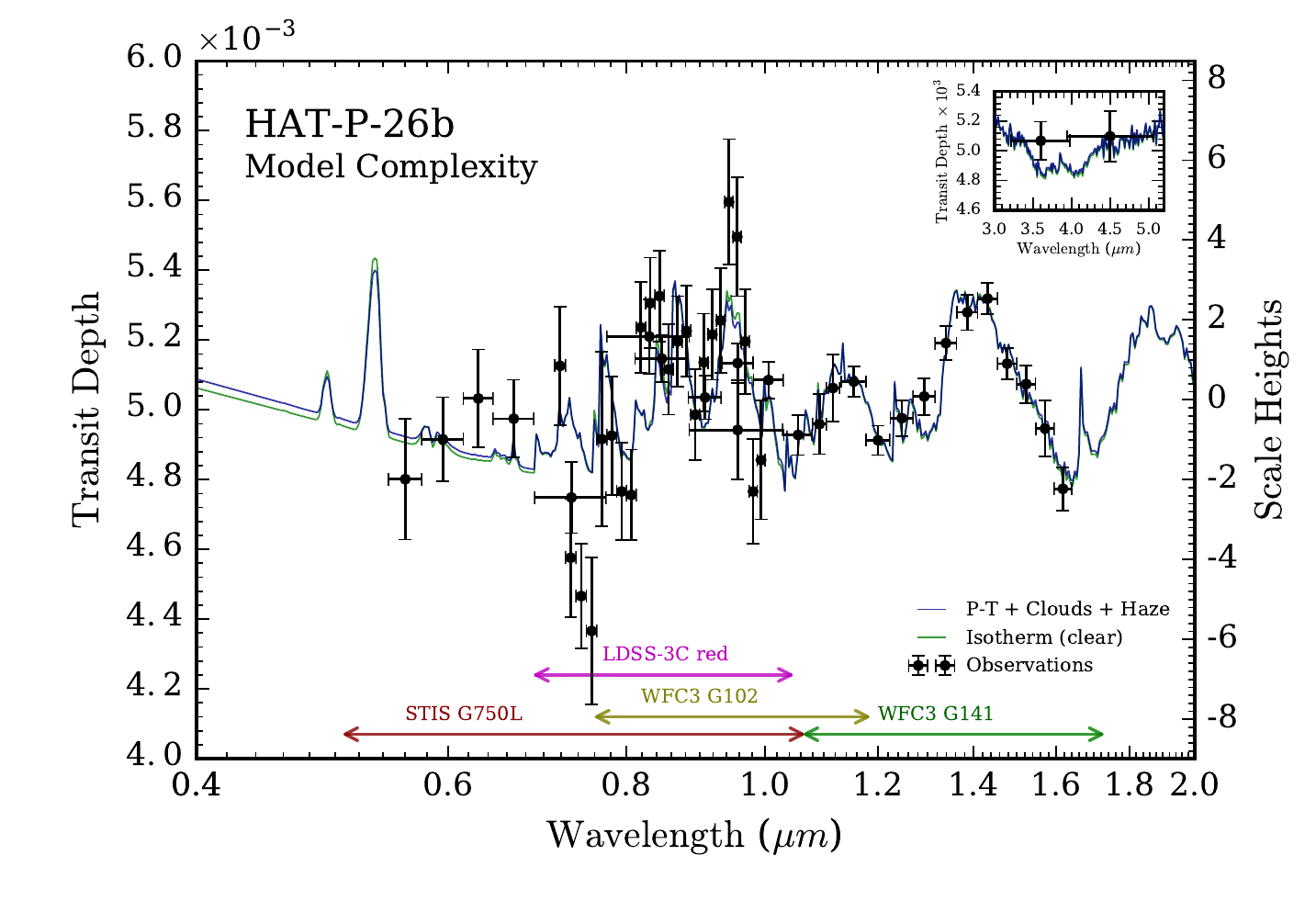}
	\caption{\textbf{The impact of non-isothermal P-T profiles and clouds on retrieved transmission spectra of HAT-P-26b}. The maximum likelihood spectra from two retrieval models are shown: (i) non-isothermal P-T profiles including clouds and hazes (blue); (ii) isothermal, clear, atmosphere (green). Both spectra are plotted at $R=300$. The observed transit depths are shown by black circles, with corresponding instrument modes and spectral ranges shown. A relative offset of +125ppm has been applied to the LDSS-3C red observations, based on the median retrieved $\delta_{\mathrm{rel}}$. The best-fitting spectra from the two models are nearly identical, indicating that the additional complexity of non-isothermal temperature structures and clouds / hazes are not required to explain the observations. \textbf{Inset}: region surrounding Spitzer band-passes.}
	\vspace{-0.2cm}
    \label{fig:Spectrum_PT_cloud_vs_isotherm_clear}
\end{figure}

\begin{figure}
	\includegraphics[width=\linewidth,  trim={-1.0cm 4.0cm -1.0cm 0.2cm}]{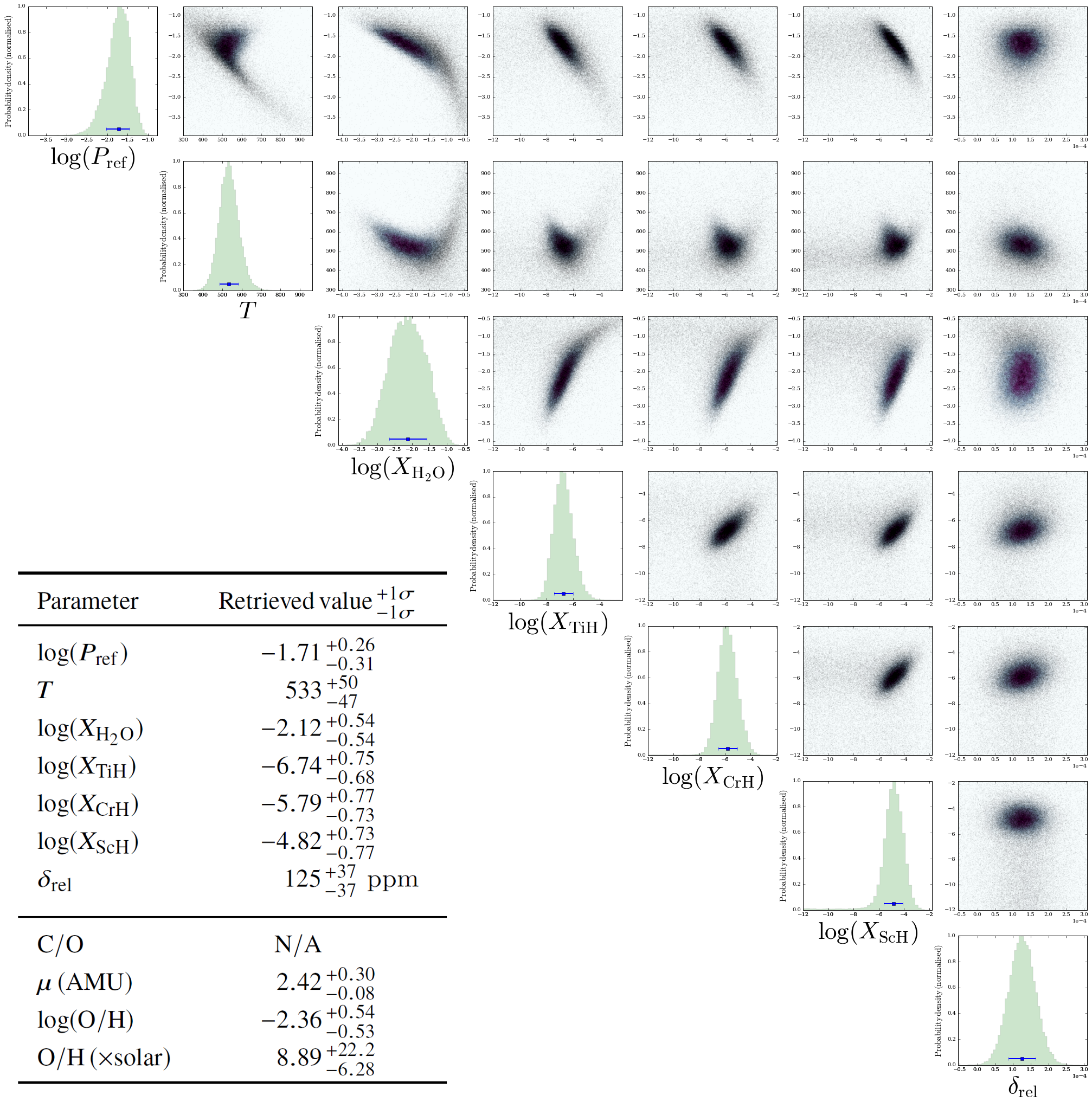} \caption{\textbf{Posterior distribution from the `minimal complexity' retrieval of HAT-P-26b}. The corner plot depicts correlations between pairs of retrieved parameters and marginalised histograms for the values of each parameter extracted by POSEIDON. The table inset summarises the statistical inferences. For each parameter, the median retrieved value and $\pm 1\sigma$ confidence levels are given. Derived atmospheric properties, as discussed in section \ref{subsubsection:derived}, are shown at the bottom of the table. C/O is given as N/A, due to it being formally zero in retrievals with no carbon-bearing species. For the posterior of the `full' retrieval (section \ref{subsection:results_spectrum}), see the supplementary online material.}
    \label{fig:posterior_isotherm}
\end{figure}



The best-fitting spectra from the `P-T + Clouds' and `Iso, Clear Skies' retrievals are compared in Figure \ref{fig:Spectrum_PT_cloud_vs_isotherm_clear}. The fits are nearly indistinguishable, resulting in similar values of $\chi^2$, despite the latter possessing 9 less free parameters. It is this reduced number of free parameters that results in the isothermal clear model possessing the lowest reduced chi-square (1.75) of this model comparison. We thus see that the small temperature gradient inferred in section \ref{subsection:results_PT} only yields a minimal difference in the fit quality when contrasted with purely isothermal models. For comparison, we show the posterior distribution from the isothermal, clear skies, retrieval in Figure \ref{fig:posterior_isotherm}, which is the `minimal complexity' model that can explain the present observations. Despite having 23 less free parameters than the `full' model, the parameter inferences are consistent. Nevertheless, we caution against drawing inferences from such a simple model, for reasons discussed in section \ref{subsection:statistics}. In totality, this analysis demonstrates that the additional complexity of non-isothermal temperature structure, clouds, and hazes are not necessary in order to explain current observations of HAT-P-26b. This conclusion holds whether or not ground-based observations are included (see appendix \ref{section:Without_ground_data}). The results from this complexity analysis are employed in the next section to explore predictions for the JWST GTO campaigns for HAT-P-26b.

\section{Predictions for JWST observations of HAT-P-26b's atmosphere} \label{section:JWST}

Transmission spectra of HAT-P-26b are expected to be obtained during the JWST Cycle 1 Guaranteed Time Observations (GTO). Multiple instrument modes from NIRISS, NIRCam, NIRSpec, and MIRI will produce a complete transmission spectrum from 0.6-11 $\micron \,$ -- for a detailed overview of JWST's spectroscopic modes in the context of transiting exoplanets, see \citet{Beichman2014}. In this section, we build upon the atmospheric inferences previously presented to predict the constraining power of such an intensive observational campaign for a warm exo-Neptune.

Previous studies have assessed the ability of retrievals on simulated JWST spectra to constrain exoplanetary atmospheres. \citet{Barstow2015} and \citet{Greene2016} considered transmission and emission spectra of archetypal exoplanets ranging from hot Jupiters to super-Earths, generated synthetic JWST observations using various combinations of instrument modes, and conducted retrievals to extract the atmospheric state of each planet. \citet{Greene2016} concluded that C/O and metallicity constraints of 0.2 dex and 0.5 dex, respectively, are anticipated from transmission spectra with a large spectral baseline. Information content analyses of JWST modes have similarly concluded that a wide spectral range, such as provided by NIRISS SOSS and NIRSpec G395H, have the greatest constraining power \citep{Batalha2017a, Howe2017}. We therefore expect the 0.6-11 $\micron$ JWST transmission spectrum of HAT-P-26b to be rich in information. Most recently, \citet{Schlawin2018} considered transmission spectra retrievals of three representative JWST GTO targets, including HAT-P-26b, predicting C/O and metallicity constraints of $\approx$ 0.1 dex.

A range of simplifying assumptions have been employed in transmission spectra retrievals. Two common assumptions are isothermal temperature profiles and chemical equilibrium. We examine each of these in turn. Firstly, in the presence of a large temperature gradient across the pressure range probed by observations, the shapes of absorption features are altered when compared to the pure isotherm case. This is due to the underlying temperature dependence of molecular cross sections, manifesting through absorption feature maxima probing higher altitudes (cooler temperatures) than absorption minima (hotter temperatures). The assumption of an isothermal temperature structure can therefore bias abundances retrieved from JWST spectra of hot Jupiters by an order of magnitude \citep{Rocchetto2016}, demonstrating that high-quality transmission spectra contain information on terminator temperature profiles \citep{Barstow2013,Barstow2015}. Secondly, some retrieval analyses elect to impose chemical equilibrium (sometimes referred to as `chemically consistent' retrieval) to retrieve C/O ratios and metallicities \citep[e.g.][]{Schlawin2018}. The motivation behind this approach is two-fold: (i) to reduce the number of free parameters (from one parameter per retrieved species to C/O and metallicity alone); and (ii) to specify uninformative priors on the C/O and metallicity parameters \citep[see][for a discussion]{Line2013}. The main disadvantage of this approach is its inability to capture disequilibrium chemistry, which can result from processes such as vertical mixing and photochemistry \citep{MacDonald2017a}. In what follows, our analysis relaxes these two assumptions.

Our present analysis contributes two unique findings to investigations of spectral retrievals in the era of JWST. Firstly, we conduct the first investigation into the ability of JWST to detect and constrain metal hydride chemistry. Secondly, we utilise nested Bayesian model comparisons to predict detection significances of key molecules. All reported constraints are marginalised over non-isothermal temperature structures, the influence of clouds, and disequilibrium chemistry. This minimal assumption approach, combined with realistic GTO observations, demonstrates the full potential of JWST to constrain exoplanetary atmospheres.

This section is structured as follows. We outline the GTO campaigns of HAT-P-26b in section \ref{subsection:JWST_campaigns}. A reference atmospheric model and resulting transmission spectrum are described in section \ref{subsection:JWST_reference_model}. Synthetic JWST observations for this model are created in section \ref{subsection:JWST_synthetic_observations}. We discuss our retrieval methodology for JWST observations in section \ref{subsection:JWST_retrieval_methods}. Finally, in section \ref{subsection:JWST_results}, we present the results of a full retrieval analysis on the synthetic JWST observations of HAT-P-26b.

\subsection{JWST GTO campaigns for HAT-P-26b} \label{subsection:JWST_campaigns}

\begin{table}
\ra{1.3}
\caption{JWST observations planned for HAT-P-26b}
\begin{tabular*}{\columnwidth}{l@{\extracolsep{\fill}} cccccl@{}}\toprule
Mode & GTO $\#$ & $\lambda \, \, (\micron)$ & $n$ & $t_{\mathrm{exposure}}$  \\ \midrule
NIRISS SOSS & $\# \, 1312$ & $0.64-1.00$ & $5$ & $6.94$ hr \\
`` " & `` " & $0.84-2.81$ & `` " & `` " \\
NIRCam F322W2 & $\# \, 1185$ & $2.41-4.05$ & $9$ & $7.50$ hr \\
NIRSpec G395H & $\# \, 1312$ & $2.89-5.18$ & $37$ & $6.94$ hr \\
NIRCam F444W & $\# \, 1185$ & $3.86-5.00$ & $20$ & $7.51$ hr \\
MIRI LRS & $\# \, 1177$ & $5.02-11.00$ & $90$ & $7.53$ hr \\
\bottomrule
\vspace{0.1pt}
\end{tabular*}
$\textbf{Notes}:$ NIRISS rows correspond to 2nd and 1st order spectra, respectively. Wavelength ranges are clipped where transit depth uncertainties exceed 1000 ppm (e.g. MIRI LRS beyond 11 $\micron$). $n$ = number of groups per integration. Exposure times include the transit duration ($t_{14}$ = 2.455 hr) and out of transit baseline. The total number of transits across all instrument modes is 5.
\label{table:GTO_observations}
\end{table}

HAT-P-26b will be observed during three accepted JWST GTO campaigns. The instrument modes due to be employed, and corresponding wavelengths ranges, are summarised in Table \ref{table:GTO_observations}. Observations with NIRISS in the single object slitless spectroscopy (SOSS) mode and the NIRSpec G395H grism (GTO \#1312\footnote[1]{\href{https://www.stsci.edu/jwst/phase2-public/1312.pdf}{https://www.stsci.edu/jwst/phase2-public/1312.pdf}}, PI Nikole Lewis) will provide a nearly complete transmission spectrum from 0.6-5.2 $\micron$. The NIRCam grism offers two modes, the F322W2 and F444W filters (GTO \#1185\footnote[2]{\href{https://www.stsci.edu/jwst/phase2-public/1185.pdf}{https://www.stsci.edu/jwst/phase2-public/1185.pdf}}, PI Thomas Greene), providing additional coverage from 2.4-5.0 $\micron$. Finally, a single transit observed with MIRI's low-resolution spectrometer (LRS) mode (GTO \#1177\footnote[3]{\href{https://www.stsci.edu/jwst/phase2-public/1177.pdf}{https://www.stsci.edu/jwst/phase2-public/1177.pdf}}, PI Thomas Greene) extends the wavelength range beyond 5 $\micron$ up to 11 $\micron$. As simultaneous observations in different modes is not possible, a total of 5 transits will be observed. By combining these observations, a complete transmission spectrum from 0.6-11 $\micron$ will be obtained.

\subsection{Reference model atmosphere of HAT-P-26b} \label{subsection:JWST_reference_model}

To simulate JWST observations of HAT-P-26b, a model atmosphere must be created. A reference transmission spectrum is then constructed by solving the equation of radiative transfer through the atmosphere in primary transit geometry. We first describe the composition and atmospheric properties of our chosen reference model in section \ref{subsubsection:JWST_reference_model_atmosphere}, before breaking down the resulting transmission spectrum into its constituent chemical signatures in section \ref{subsubsection:JWST_reference_model_signatures}.

\subsubsection{Model atmosphere} \label{subsubsection:JWST_reference_model_atmosphere}

Our reference model atmosphere is constructed to be consistent with inferences from current observations (section \ref{section:results}) and expectations from chemical models of exo-Neptune atmospheres \citep[see][for a review]{Madhusudhan2016}. For the chemical species inferred in section \ref{subsection:results_composition}, we select mixing ratios representative of the modal values in Figure \ref{fig:abundances}: $\mathrm{log}(X_{\mathrm{H_2 O}}) = -1.83$, $\mathrm{log}(X_{\mathrm{TiH}}) = -6.2$, $\mathrm{log}(X_{\mathrm{CrH}}) = -5.7$, and $\mathrm{log}(X_{\mathrm{ScH}}) = -4.5$. For molecules with upper limits only, we initially select representative equilibrium mixing ratios for a 900 K exo-Neptune from \citet{Moses2013} (their Figure 11), corresponding to a metallicity near to our derived value ($\sim 20 \, \times$ solar, Figure \ref{fig:derived}). The mixing ratios were then adjusted as necessary to ensure consistency with our derived upper limits on both the mixing ratios and C/O. The resulting chosen values are as follows: $\mathrm{log}(X_{\mathrm{CH_4}}) = -5.5$, $\mathrm{log}(X_{\mathrm{CO}}) = -2.7$, $\mathrm{log}(X_{\mathrm{CO_2}}) = -4.5$, $\mathrm{log}(X_{\mathrm{NH_3}}) = -6.0$, and $\mathrm{log}(X_{\mathrm{N_2}}) = -3.0$. It follows that this reference atmosphere has a metallicity of $20.1 \, \times$ solar, C/O = 0.12, and $\mu$ = 2.62.

Similarly, the temperature structure and cloud properties are chosen with our previous inferences in mind. The complexity analysis in section \ref{subsection:results_model_complexity} identified a slight preference for models with non-isothermal P-T profiles. We thus adopt a P-T profile with $T_{\mathrm{1 \, mbar}} =560$ K and a temperature gradient of $\sim$ 80 K across the infrared photosphere -- consistent with the median retrieved P-T profile in Figure~\ref{fig:PT}. $P_{\mathrm{ref}}$ is set to 10 mbar (consistent with Figure~\ref{fig:posterior_isotherm}). As neither our retrieved posterior or complexity analysis favours the inclusion of clouds or hazes, the reference model is chosen to have clear skies. We do, however, include clouds and hazes as parameters during our retrieval analysis in section \ref{subsection:JWST_results}, in order to ensure that all reported constraints include any degeneracies induced by these parameters.

\subsubsection{Transmission spectrum and chemical signatures} \label{subsubsection:JWST_reference_model_signatures}

\begin{figure}
	\includegraphics[width=\linewidth,  trim={1.0cm 1.2cm 0.4cm 0.4cm}]{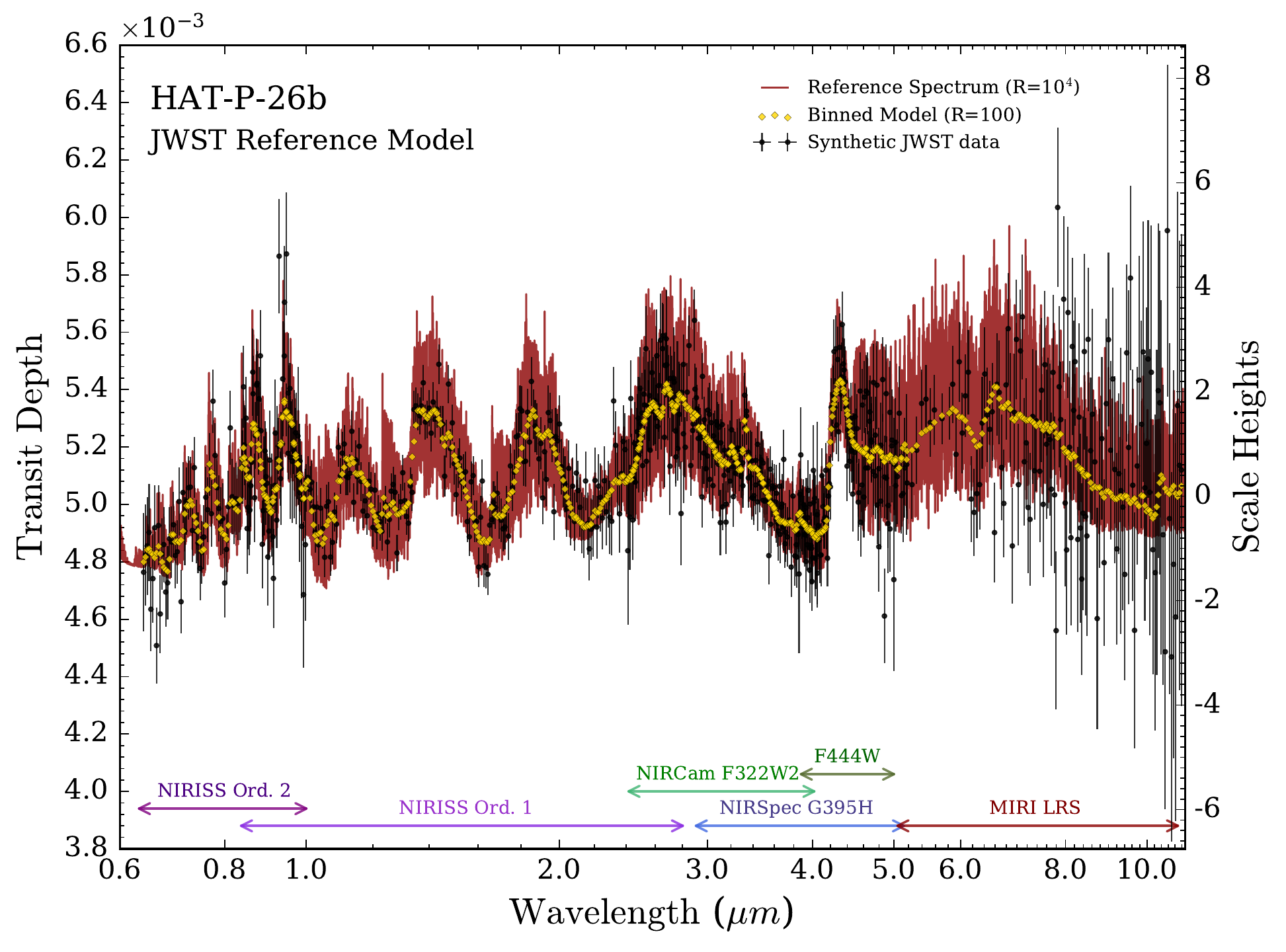}
	\caption{\textbf{Reference transmission spectrum of HAT-P-26b and synthetic JWST observations}. A reference model atmosphere, consistent with the findings in section \ref{section:results}, results in the reference transmission spectrum, computed at a spectral resolution of $R=10^{4}$ (red). Simulated JWST observations, generated using PandExo \citep{Batalha2017b}, result in the black synthetic data points. The instrument modes and spectral ranges, shown at the bottom, correspond to the GTO campaigns for HAT-P-26b. The gold diamonds show the reference spectrum binned to the same resolution as the synthetic observations ($R=100$).}
	\vspace{-0.4cm}
    \label{fig:JWST_reference_spectrum}
\end{figure}

With our reference atmosphere in hand, we now turn to discus the corresponding transmission spectrum. The equation of radiative transfer is solved as outlined in section  \ref{subsection:radiative_transfer}, with one exception: the model wavelength range is extended to 0.4-12 $\micron$ and the computed spectral resolution is increased to $R=10^4$. The former is required to simulate MIRI observations beyond 5 $\micron$, whilst the latter ensures the model resolution exceeds the maximum resolving power of JWST's instrument modes ($R_{\mathrm{max}} \approx 3,500$ for NIRSpec). The resulting transmission spectrum is shown in Figure \ref{fig:JWST_reference_spectrum}. For a more direct comparison with the observable spectrum, we also convolve this spectrum with the PSF of each instrument mode and integrate over the corresponding sensitivity functions to produce binned model points at $R=100$.

\begin{figure*}
	\includegraphics[width=\linewidth,  trim={-1.33cm 0.7cm -1.33cm 0.4cm}]{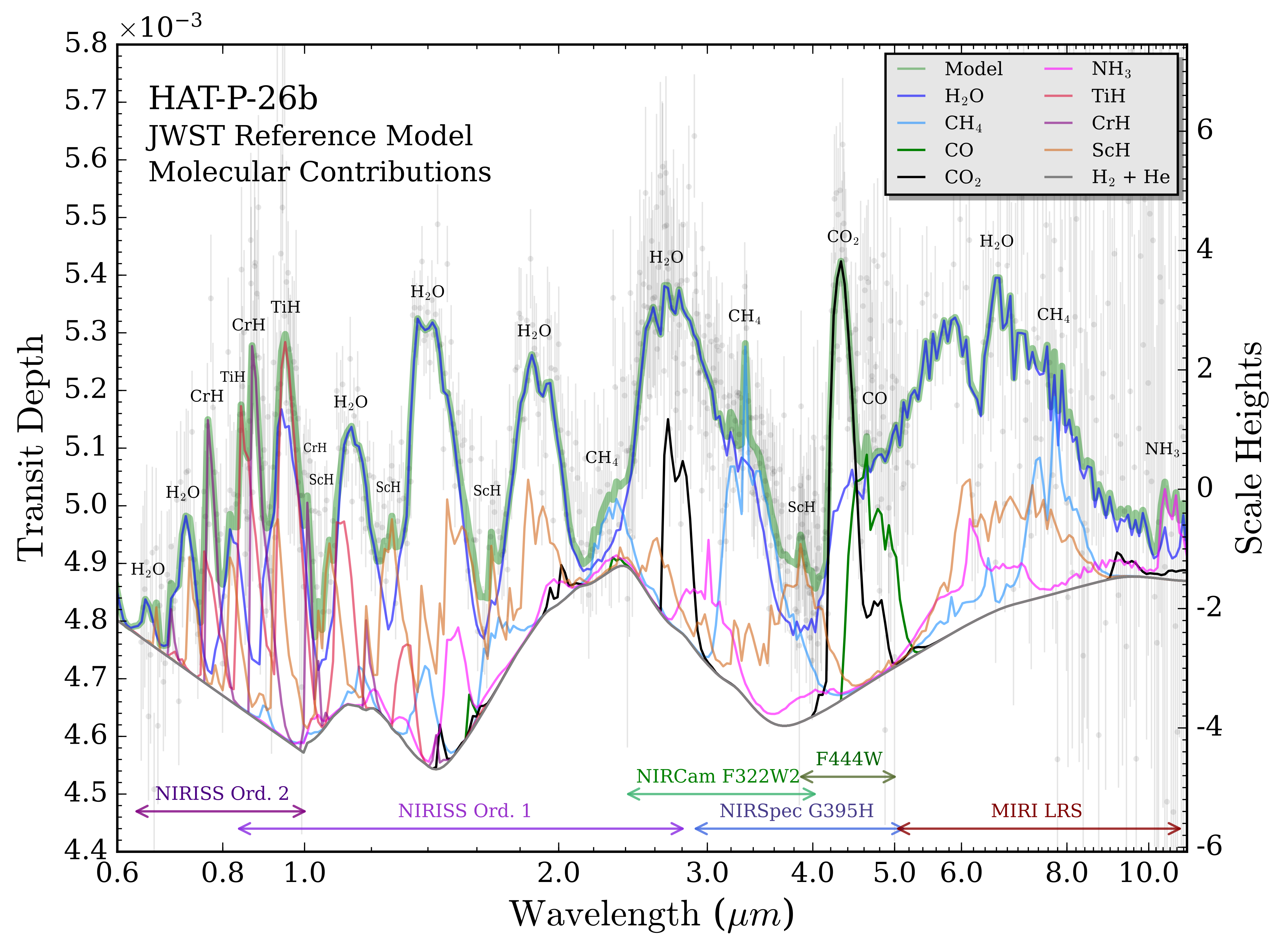}
	\caption{\textbf{Molecular contributions to the reference transmission spectrum of HAT-P-26b}. The green shading indicates the reference spectrum of HAT-P-26b, calculated at $R=100$ for comparison with the simulated JWST observations. The grey line shows the spectral continuum from H$_2$ and He alone. Other lines display contributions of each molecule above the H$_2$+He continuum. Prominent absorption features are labelled. Note that N$_2$ has no notable absorption features (due to its lack of an electric dipole moment), instead indirectly influencing the spectrum through minor contributions to the mean molecular weight and Rayleigh slope. Simulated JWST observations, generated by PandExo \citep{Batalha2017b}, are overplotted as transparent circles with error bars for comparison to the amplitudes of spectral features. The corresponding instrument modes and spectral ranges are shown at the bottom.}
    \label{fig:JWST_spectral_contributions}
\end{figure*}

The contributions of each molecule to the reference transmission spectrum are shown in Figure \ref{fig:JWST_spectral_contributions}. By comparing the absorption feature amplitudes of each chemical species to typical noise levels of the synthetic observations (see section \ref{subsection:JWST_synthetic_observations}), one can gain qualitative intuition for which species will be detectable with JWST. We rigorously quantify detection significances in section \ref{subsubsection:JWST_results_composition}, but such an exercise is nevertheless instructive prior to a full retrieval analysis. As the increased wavelength range and coverage provided by JWST observations introduces signatures of additional molecules than those discussed in section \ref{subsubsection:signatures} and Figure \ref{fig:spectral_signatures}, we now turn to examine the anticipated prominence of each molecular species.

The infrared transmission spectrum beyond 1 $\micron$ is primarily shaped by H$_2$O absorption, with smaller contributions from other molecules containing carbon, oxygen, nitrogen, and hydrogen. H$_2$O is prominent across all instrument modes, indicating an extremely strong anticipated detection ($>$ 10$\sigma$). CO$_2$ possesses the strongest absorption feature in our spectrum, occurring around 4.3 $\micron$. As this region is sampled by both NIRCam F444W and NIRSpec G395H, we similarly expect a strong detection (e.g. $>$ 10 simulated observations are $>$ 5$\sigma$ devient from the contribution of H$_2$O alone). CH$_4$ is notable around 2.3 $\micron$ and 3.4 $\micron$, with a minor contribution around 8 $\micron$. Even at the relatively low chosen abundance ($10^{-5.5}$), we anticipate a detection of CH$_4$ to arise from combining NIRISS, NIRCam F322W2, and NIRSpec G395H observations. Detections of CO are expected to be challenging, as it only contributes around 4.6 $\micron$. NH$_3$ is not expected to be detectable at the chosen abundance ($10^{-6}$), as its only prominent signature (around 10.5 $\micron$) is buried in the high-noise long-wavelength MIRI data. We note that higher NH$_3$ abundances raise the possibility of detections by NIRISS in the strong 2.2 $\micron$ feature \citep{MacDonald2017b}.  

Signatures of metal hydrides largely dominate sub-micron wavelengths. Multiple sharp TiH and CrH features are probed by NIRISS SOSS, with the region from 0.84-1.0 $\micron$, where the 1st and 2nd NIRISS orders overlap, coinciding with the strongest features of TiH and CrH. The additional short wavelength coverage of the 2nd order down to 0.64 $\micron$ further enhances the ability to assess the presence of TiH and CrH. We thus expect strong detections of these two metal hydrides. The utility of NIRISS to constrain the composition of exoplanetary atmospheres at $\lambda < 1 \micron$ merits particular focus as, to our knowledge, this capability has not been studied in previous retrieval analyses of JWST spectra. However, we note that probing metal hydrides in this manner is more difficult for stars with J magnitude $<$ 8.5 (brighter targets saturate the SUBSTRIP256 subarray required for 2nd order observations). Absorption due to ScH, conversely, is spread more evenly over the infrared, occurring in three regions of minimal H$_2$O absorption observable by NIRISS and around 4.0 $\micron$ accessible to both NIRCam filters and NIRSpec G395H. This latter signature, covered by three instrument modes, will provide a powerful diagnostic of the presence (or absence) of ScH. With the key signatures of potentially observable molecules in mind, we now address the creation of synthetic observations for this spectrum.

\subsection{Synthetic JWST observations of HAT-P-26b} \label{subsection:JWST_synthetic_observations}

Synthetic JWST observations are generated using PandExo \citep{Batalha2017b}. The assumed stellar spectrum for HAT-P-26 is interpolated from the Phoenix Stellar Atlas \citep{Husser2013} to $T_{\mathrm{eff}}$ = 5080 K, [Fe/H] = -0.04, $\mathrm{log}(g)$ = 4.56 (cgs), and normalised to J = 10.08. Given this stellar spectrum, and the reference planetary transmission spectrum from section \ref{subsection:JWST_reference_model}, in transit and out of transit fluxes are computed. Realistic noise levels are calculated for each instrument mode discussed in section \ref{subsection:JWST_campaigns}, using exposure times and numbers of groups per integration as specified in each GTO proposal and repeated in Table \ref{table:GTO_observations}. A saturation limit of 80\% full well is assumed throughout. Noise floors are set at 20 ppm, 30 ppm, 30 ppm, and 50 ppm for NIRISS, NIRCam, NIRSpec, and MIRI, respectively \citep{Beichman2014,Greene2016}. The resulting transit depths and uncertainties are binned to a common spectral resolution of $R=100$, clipping wavelength ranges where falling sensitivity curves for a given mode dramatically increase transit depth uncertainties (e.g. MIRI LRS uncertainties beyond 11 $\micron$ exceed 1000 ppm). The final set of synthetic observations consists of 714 data points, with a mean precision of 120 ppm for 0.6-5.0 $\micron$ and 300 ppm beyond 5 $\micron$. The synthetic observations are shown in Figures \ref{fig:JWST_reference_spectrum} and \ref{fig:JWST_spectral_contributions}.

\begin{figure*}
	\includegraphics[width=\linewidth,  trim={-2.5cm 0.8cm -2.5cm 0.6cm}]{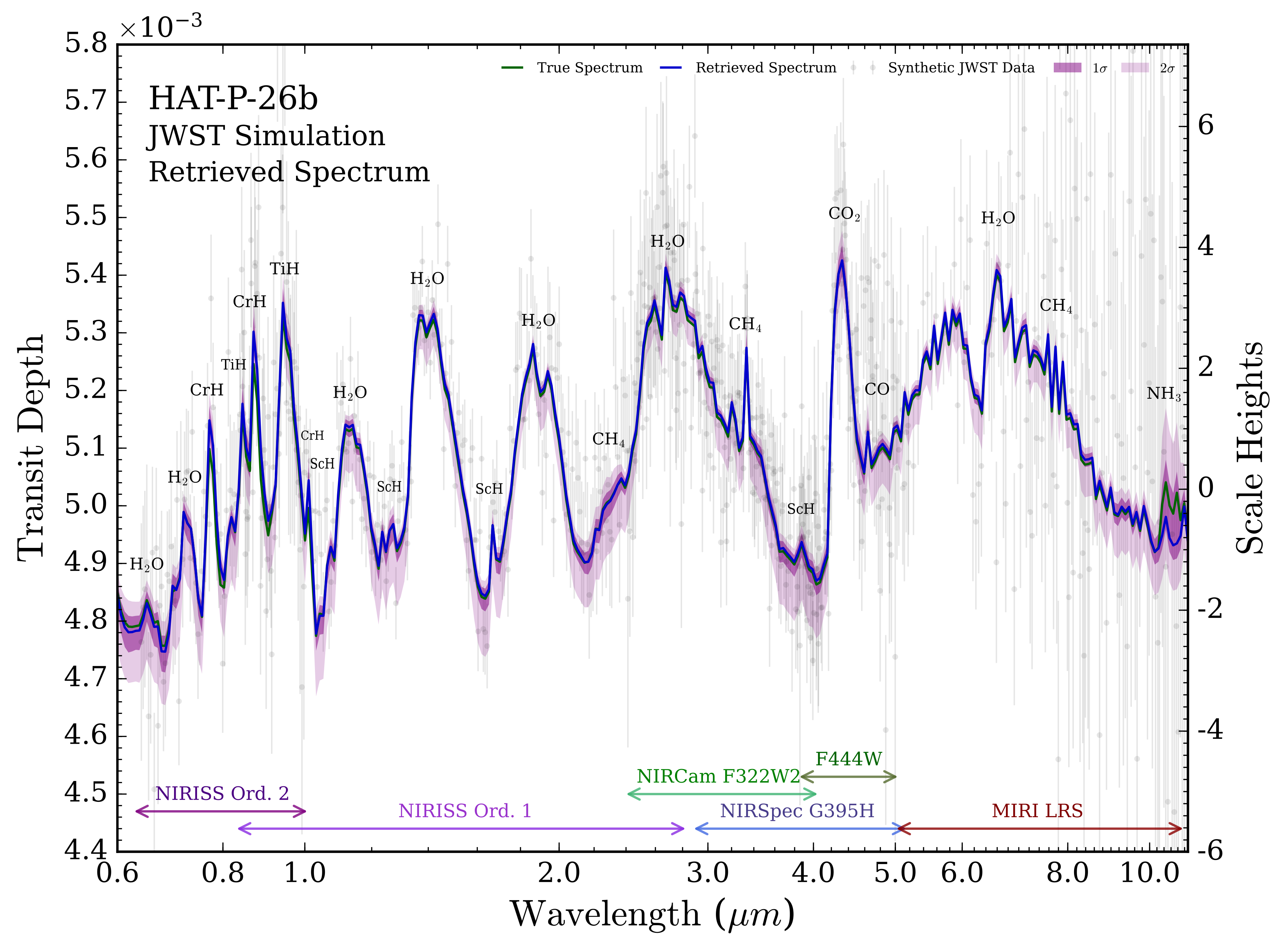}
    \caption{\textbf{Retrieved transmission spectrum from simulated JWST observations of HAT-P-26b}. Simulated JWST observations (transparent circles with error bars), with corresponding instrument modes and spectral ranges at the bottom, undergo a spectral retrieval by POSEIDON. The true input transmission spectrum (green) compares well with the median retrieved spectrum (blue), both plotted at the same spectral resolution as the synthetic observations ($R = 100$). 1$\sigma$ and 2$\sigma$ confidence regions (purple) are derived from 30,000 random posterior samples. Prominent absorption features due to specific molecules are labelled.}
    \label{fig:JWST_spectrum_retrieved}
\end{figure*}

\subsection{Retrieval methodology for JWST observations} \label{subsection:JWST_retrieval_methods}

A full retrieval analysis can now be conducted on the synthetic JWST observations. The atmospheric parameterisation is broadly as discussed in section \ref{subsection:parameters}. We parameterise the following molecules: H$_2$O, CH$_4$, NH$_3$, CO, CO$_2$, AlO, TiH, CrH, and ScH. This set contains the molecules used to generate the reference spectrum (apart from N$_2$, which is negligible at the assumed abundance). AlO is included to ensure the degeneracy with ScH, discussed in section \ref{subsubsection:detections}, is factored into detection significances. Clouds and hazes are parameterised as in Table \ref{table:priors}. The P-T profile parameterisation is modified to replace $P_{\mathrm{ref}}$ with a new parameter, $R_{\mathrm{p, \, 10 mbar}}$, as we found that retrievals with this synthetic dataset are precise enough to encounter artefacts due to the discrete atmospheric pressure grid when $P_{\mathrm{ref}}$ is a free parameter. The reference model condition that $P_{\mathrm{ref}}$ = 10 mbar is equivalent to $R_{\mathrm{p, \, 10 mbar}}$ = $R_{\mathrm{p, \, obs}} = 0.544 R_{\mathrm{J}}$. A uniform prior is prescribed for $R_{\mathrm{p, \, 10 mbar}}$ between 80\% and 120\% of the true value. In total, this amounts to 20 free parameters.

A total of 13 atmospheric retrievals were conducted. This consists of a reference retrieval, 9 retrievals with each parameterised molecule removed in turn, a retrieval without H$_2$+He, a retrieval without metal hydrides, and a retrieval assuming an isothermal temperature structure. Spectra are generated from 0.6-11.0 $\micron$ at $R=5000$, at $> 5 \times 10^{6}$ parameter combinations for each retrieval. These nested retrievals allow computation of molecular detection significances and an assessment of biases caused by an isothermal atmosphere. We now present the results of this analysis.

\subsection{Results: characterising an exo-Neptune atmosphere with JWST} \label{subsection:JWST_results}

Here we demonstrate the potential of JWST to characterise an exo-Neptune atmosphere. We first discuss the ability to recover the reference transmission spectrum and temperature structure of HAT-P-26b in sections \ref{subsubsection:JWST_results_spectrum} and \ref{subsubsection:JWST_results_PT}, respectively. We then offer predicted detection significances and abundance constraints for each molecular species in section \ref{subsubsection:JWST_results_composition}, including an elucidation of biases caused by assuming an isothermal temperature structure. Finally, in section \ref{subsubsection:JWST_results_derived}, we predict the ability to constrain the metallicity, mean molecular weight, and C/O of HAT-P-26b's atmosphere following the scheduled JWST observations.

\begin{figure*}
	\includegraphics[width=\linewidth,  trim={-4.2cm 7.0cm -4.2cm 4.0cm}]{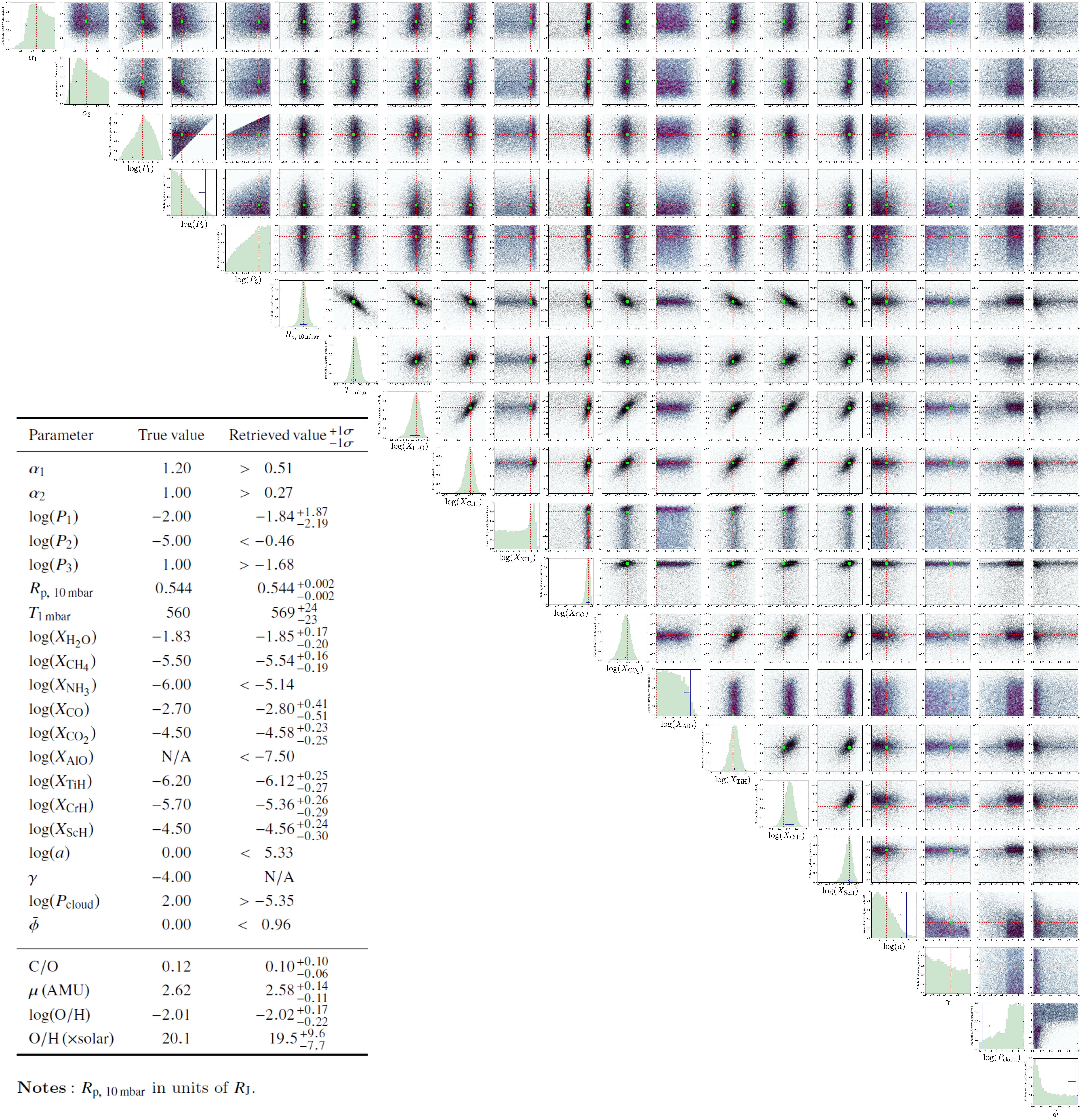} \caption{\textbf{Posterior distribution from the `full' retrieval of synthetic JWST observations for HAT-P-26b}. The corner plot depicts correlations between pairs of retrieved parameters and marginalised histograms for each parameter extracted by POSEIDON. Red dashed lines and green squares indicate the true parameter values used to generate the data. The table inset compares true values with retrieved inferences. Where parameters have clear upper and lower bounds, the median retrieved values and $\pm 1\sigma$ confidence levels are given. Otherwise, parameter constraints are expressed as $2\sigma$ upper or lower bounds. N/A indicates redundant parameters ($\gamma$ due to the unconstrained distribution). Derived atmospheric properties, discussed in section \ref{subsubsection:JWST_results_derived}, are shown at the bottom of the table.}
    \label{fig:JWST_posterior_full}
\end{figure*}

\subsubsection{Retrieved transmission spectrum} \label{subsubsection:JWST_results_spectrum}

Figure \ref{fig:JWST_spectrum_retrieved} compares the true reference transmission spectrum of HAT-P-26b to that retrieved by POSEIDON. We see that the median retrieved spectrum generally lies within 1$\sigma$ of the true spectrum over the full 0.6-11.0 $\micron$ range. The one exception is the CrH features, for which the retrieved spectrum overestimates the amplitude by $\sim$ 2$\sigma$. This is due to an excess of elevated data points around the CrH feature at 0.88 $\micron$ for this particular noise instance. We thus expect the retrieved CrH abundance to be higher than the true value -- a point we revisit in section \ref{subsubsection:JWST_results_composition}. Despite this, the minimum reduced chi-square indicates an excellent fit ($\chi_{r, \mathrm{min}}^{2}$ = 1.04 for 694 degrees of freedom). Overall, the retrieval correctly identifies the molecules responsible for the main absorption features. Remarkably, there is even a hint of the NH$_3$ feature around 10.5 $\micron$, though the 1$\sigma$ contour remains consistent with models without NH$_3$ (matching the qualitative picture provided in section \ref{subsubsection:JWST_reference_model_signatures}).

The posterior distribution corresponding to the retrieved transmission spectrum is shown in Figure \ref{fig:JWST_posterior_full}. Most parameters are retrieved within 1$\sigma$ of their true value (the one exception being the CrH abundance, as previously discussed). Molecules significantly contributing to the spectrum are well-constrained. In particular, the ScH-AlO degeneracy is broken, with the synthetic JWST data sufficient to set an upper bound on AlO. We note that the retrieval identifies two equivalent solutions to explain the clear atmosphere: deep clouds with variable cloud coverage or high-altitude clouds with zero cloud coverage (Figure \ref{fig:JWST_posterior_full}, lower two histograms). Contrasting this posterior with that from current observations (supplementary material), new constraints on the P-T profile parameters are now evident (e.g. the $\alpha_2$ - $\mathrm{log}(P_1)$ plane). The implications of these enhanced constraints for the retrieved P-T profile is the next focus.

\subsubsection{Temperature structure} \label{subsubsection:JWST_results_PT}

\begin{figure}
	\includegraphics[width=0.95\linewidth,  trim={0.0cm 0.4cm 0.0cm 0.4cm}]{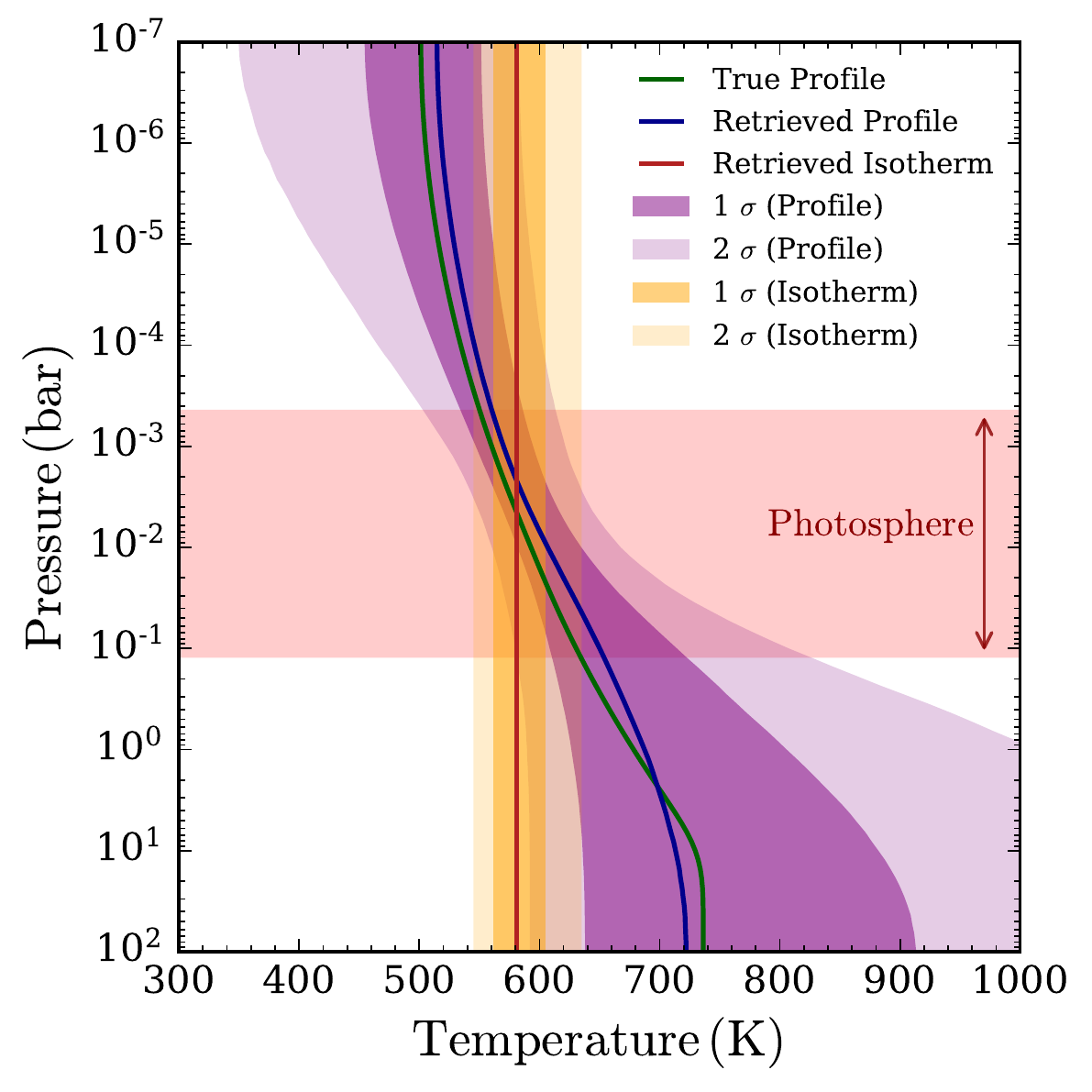}
	\caption{\textbf{Retrieved terminator temperature structure from simulated JWST observations of HAT-P-26b}. The true input profile (green) is compared against median profiles from two independent retrievals: (i) a retrieval with a flexible P-T profile (blue), and (ii) an isothermal temperature structure (red). Corresponding 1$\sigma$ and 2$\sigma$ confidence regions (purple for the P-T profile, orange for the isotherm) are derived from 30,000 random posterior samples. The isotherm agrees with the true profile across a decade in pressure around $\sim$ 10 mbar, whilst the P-T profile is consistent with the true profile over the entire atmosphere. The $\tau \leq 1$ infrared photosphere probed by the observations is shaded in red. The pressures probed range from low pressures coinciding with absorption features to deep regions probing the continuum.}
    \label{fig:JWST_PT}
    \vspace{-0.4cm}
\end{figure}

We now assess the ability to retrieve terminator temperature structures from JWST observations of an exo-Neptune. Figure \ref{fig:JWST_PT} compares the true P-T profile of the reference model atmosphere with profiles obtained from two retrievals: (i) the `full' retrieval with the 6 parameter P-T profile of \citet{Madhusudhan2009}, and (ii) a single parameter isothermal model. Both model profiles are supplemented by the $R_{\mathrm{p, \, 10 mbar}}$ parameter discussed in section \ref{subsection:JWST_retrieval_methods}. We see that the flexible P-T profile correctly matches the true profile within 1$\sigma$ over the entire atmosphere, whilst the isotherm is only consistent with the true profile from $\sim$2 - 30 mbar. The Bayesian evidence supports the interpretation favouring a non-isothermal profile, albeit marginally ($\mathcal{B}_{\mathrm{ij}}$ = 1.3). The retrieved 1 mbar temperature, $T_{\mathrm{1 mbar}}$ = $569^{+24}_{-23}$ K, agrees well with the true value of 560 K (cf. $T = 584^{+24}_{-19}$ K for the isothermal model). This constraint represents a 2.4 $\times$ improvement over the $\sim$ 57 K uncertainty obtained from current observations (see Figure \ref{fig:PT}). The temperature confidence intervals are tightest around 10 mbar -- corresponding to the centre of the infrared photosphere -- expanding significantly for pressures outside those directly probed by the observations ($P <$ 0.1 mbar and $P >$ 100 mbar). In totality, these results suggest that the photosphere is wide enough for the synthetic JWST observations to be sensitive to a temperature gradient as small as $\sim$ 80 K. We examine the effect of ignoring this retrievable temperature gradient on the inferred chemical abundances in the next section.

\subsubsection{Composition} \label{subsubsection:JWST_results_composition}

The wavelength coverage and spectral resolution of JWST will significantly enhance constraints on the composition of exoplanetary atmospheres. This will manifest in two primary ways: (i) detections of new chemical species, along with increased detection significances for known species; and (ii) more stringent constraints on chemical abundances than is currently possible. Here we provide quantitative predictions for the performance of JWST in these two areas for realistic GTO campaigns of an exo-Neptune atmosphere.

Our analysis predicts new chemical species will be conclusively detected ($>$ 5$\sigma$) by JWST. The Bayesian model comparison arriving at these detection significances is given in Table \ref{table:JWST_composition_models}. From highest to lowest significance, we anticipate detections of: CO$_2$ (13.1$\sigma$), TiH (8.3$\sigma$), CrH (8.0$\sigma$), CH$_4$ (6.2$\sigma$), and ScH (5.1$\sigma$). These detections join those of H$_2$+He and H$_2$O, at 25.8$\sigma$ and 28.9$\sigma$, respectively. CO falls short of a conclusive detection (3.7$\sigma$), but would still be considered a `strong' detection on the Jeffreys' scale ($\mathcal{B}_{\mathrm{ij}} > 150$). No evidence is found supporting the presence of NH$_3$ or AlO. The former finding reaffirms the qualitative expectation that the NH$_3$ abundance lies below the threshold to be extracted from the noise. The latter finding indicates the degeneracy between ScH and AlO seen in our analysis of current observations should not preclude definitive identification of ScH with JWST. Finally, we find the combination of three metal hydrides inferred from current observations should be detectable at 13.7$\sigma$ confidence by the planned JWST GTO campaigns. This provides a testable prediction to readily allow this hypothesis to be critically assessed. \newline

\begin{table}
\ra{1.3}
\caption[]{Predicted molecular detection significances from JWST}
\begin{tabular*}{\columnwidth}{l@{\extracolsep{\fill}} cccccl@{}}\toprule
$\mathrm{Model}$ & \multicolumn{1}{p{1cm}}{\centering \hspace{-0.4cm} Evidence \\ \centering $ \hspace{-0.2cm} \mathrm{ln}\left(\mathcal{Z}_{i}\right)$}  & \multicolumn{1}{p{1cm}}{\centering Best-fit \\ \centering $ \chi_{r, \mathrm{min}}^{2}$} & \multicolumn{1}{p{1.7cm}}{\centering \hspace{-0.3cm} Bayes \\ \hspace{-0.2cm} Factor \\ \centering $ \hspace{-0.2cm} \mathcal{B}_{0i}$}& \multicolumn{1}{p{1cm}}{\centering \hspace{-0.6cm} Significance \\ \centering \hspace{-0.4cm} of Ref.}\\ \midrule
\textbf{Full Chem} & $ 5371.0 $ & $ 1.04 $ & Ref. & Ref.\\
\hspace{0.2 em} No H$_2$+He & $ 5041.0 $ & $ 2.01 $ & $ 2 \times 10^{143} $ & $25.8 \sigma$ \\
\hspace{0.2 em} No H$_2$O & $ 4957.3 $ & $ 2.17 $ & $ 5 \times 10^{179} $ & $28.9 \sigma$ \\
\hspace{0.2 em} No CH$_4$ & $ 5353.7 $ & $ 1.09 $ & $ 3.5 \times 10^{7} $ & $6.2 \sigma$ \\
\hspace{0.2 em} No NH$_3$ & $ 5371.2 $ & $ 1.03 $ & $ 0.80 $ & N/A \\
\hspace{0.2 em} No CO & $ 5365.7 $ & $ 1.05 $ & $ 198 $ & $3.7 \sigma$ \\
\hspace{0.2 em} No CO$_2$ & $ 5287.4 $ & $ 1.27 $ & $ 2 \times 10^{36}$ & $13.1 \sigma$ \\
\hspace{0.2 em} No AlO & $ 5371.8 $ & $ 1.03 $ & $ 0.40 $ & N/A \\
\hspace{0.2 em} No TiH & $ 5338.7 $ & $ 1.13 $ & $ 1 \times 10^{14} $ & $8.3 \sigma$ \\
\hspace{0.2 em} No CrH & $ 5341.0 $ & $ 1.13 $ & $ 1 \times 10^{13} $ & $8.0 \sigma$ \\
\hspace{0.2 em} No ScH & $ 5360.1 $ & $ 1.06 $ & $ 6 \times 10^{4} $ & $5.1 \sigma$ \\
\hspace{0.2 em} No M-Hydrides & $ 5280.2 $ & $ 1.26 $ & $ 3 \times 10^{39} $ & $13.7 \sigma$ \\
\bottomrule
\vspace{0.05pt}
\end{tabular*}
$\textbf{Notes}:$ The `Full Chem' reference model includes chemical opacity due to H$_2$, He, H$_2$O, CH$_4$, NH$_3$, CO, CO$_2$, AlO, TiH, CrH, and ScH. The `No M-Hydrides' model has TiH, CrH, and ScH removed. The number of degrees of freedom (d.o.f), given by $N_{\mathrm{data}} - N_{\mathrm{params}}$, is 694 for the reference model ($N_{\mathrm{data}}$ = 714). $\chi_{r, \, \mathrm{min}}^{2}$ is the minimum reduced chi-square ($\chi^2$/d.o.f). The significance indicates the degree of preference for the reference model, highlighted in bold, over each alternative model. N/A indicates no (or negative) evidence ($\mathcal{B}_{\mathrm{ij}} \lesssim 1$) supporting a given molecule.
\label{table:JWST_composition_models}
\end{table}

\begin{figure*}
	\includegraphics[trim={-0.4cm 0.4cm -0.4cm 0.2cm}]{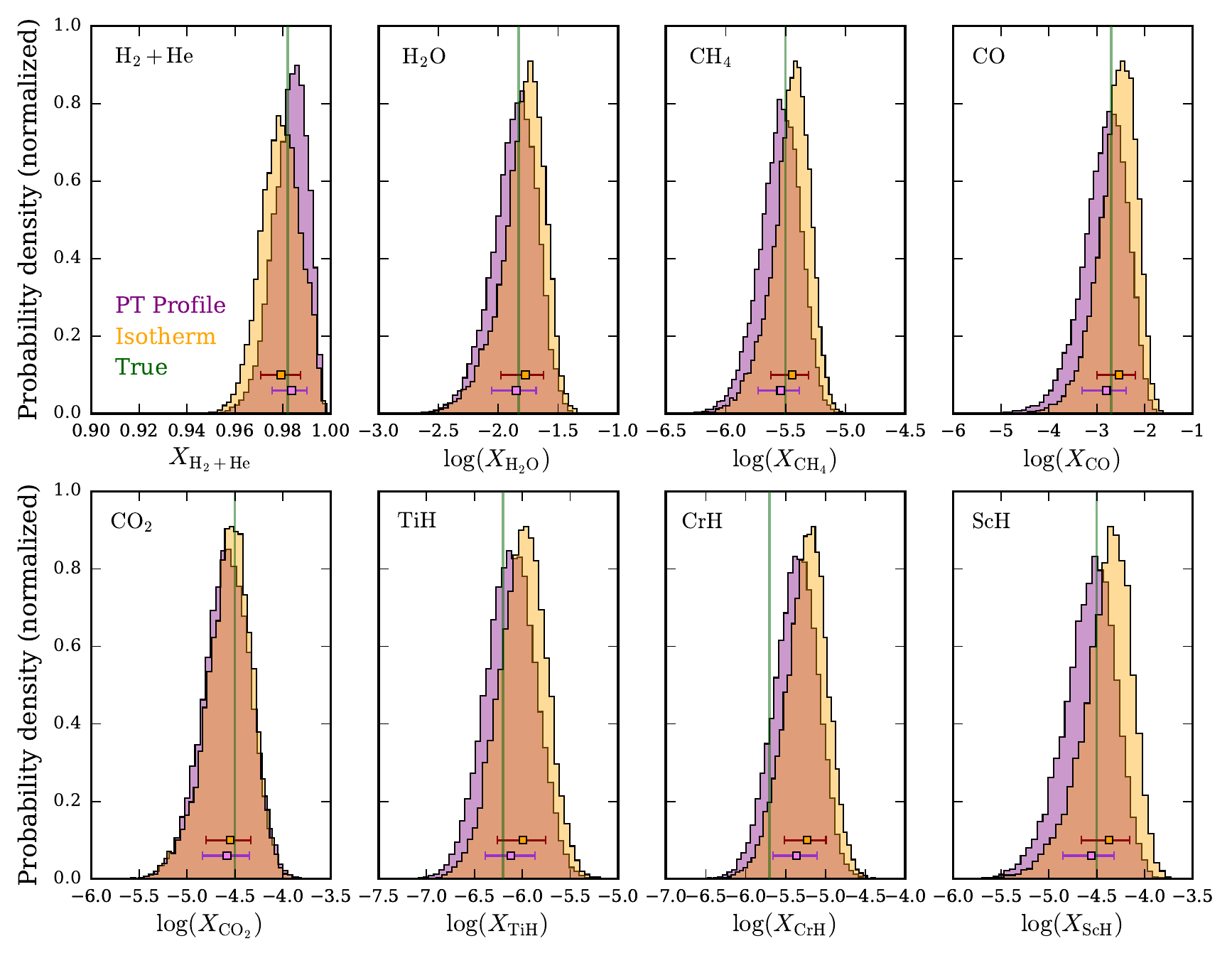}
	\caption{\textbf{Retrieved composition from synthetic JWST observations of HAT-P-26b}. Two posterior probability distributions are shown for the volume mixing ratios of each constrained chemical species: (i) a retrieval with a flexible P-T profile (purple), and (ii) a retrieval assuming an isothermal temperature structure (orange). The true value of each mixing ratio is shown by the green vertical lines. The median derived abundances and $\pm$1$\sigma$ confidence levels from each retrieval are denoted by purple and orange error bars, for the P-T profile retrieval and isothermal retrieval, respectively. The abundance of the bulk component, H$_2$+He, is given in a linear scale, whilst the secondary component (H$_2$O) and trace gases are given on a log-scale.}
    \label{fig:JWST_abundances}
\end{figure*}

We predict that JWST will be capable of constraining abundances of conclusively detected chemical species to precisions below 0.3 dex. Specifically, our retrievals of simulated JWST observations find the following precisions: 0.72\% (H$_2$+He), 0.19 dex (H$_2$O), 0.17 dex (CH$_4$), 0.46 dex (CO), 0.24 dex (CO$_2$), 0.26 dex (TiH), 0.28 dex (CrH), and 0.27 dex (ScH). This represents an improvement by a factor of $>$ 2 for the H$_2$+He and H$_2$O abundances over current observations, and $\gtrsim$ 3 for the metal hydrides. The posterior distributions for the abundances of these species are shown in Figure \ref{fig:JWST_abundances}, both for a retrieval assuming an isotherm and a retrieval accounting for non-isothermal P-T profiles. Most abundances are correctly retrieved within 1$\sigma$. The one exception is CrH, which lies just beyond 1$\sigma$ of the true value (for the P-T retrieval), due to random scatter in the particular noise instance drawn for the synthetic NIRISS observations. We do not show NH$_3$ here, despite a suggestive posterior peak around the true abundance (see Figure \ref{fig:JWST_posterior_full}), as the Bayesian model comparison established a non-detection.

Isothermal temperature assumptions bias retrieved abundances to higher values. This finding agrees with \citet{Rocchetto2016}, who found a bias of $\sim$ 1 dex for hot Jupiters. As our assumed P-T profile exhibits a shallower temperature gradient than their hot Jupiter models ($\sim$ 80 K vs. $\gtrsim$ 800 K), we find biases in the range $\sim$ 0.1-0.2 dex. This is sufficiently small to maintain consistency between our isothermal and non-isothermal posteriors. However, a parameterised P-T profile generally produces more accurate results. The bias to higher values for the retrieved abundances additionally biases the bulk atmospheric component (H$_2$+He, in this case) to lower abundances, due to the summation to unity condition. This, in turn, will bias the derived atmospheric mean molecular weight. This additional bias, though minor here due to the shallow temperature gradient, could prove more severe for high mean molecular weight atmospheres with large temperature gradients. We therefore advise caution in assuming isothermal temperature profiles when retrieving transmission spectra of hot exo-Neptunes.

\begin{figure*}
	\includegraphics[width=\linewidth,  trim={-0.4cm 0.9cm -0.4cm 0.5cm}]{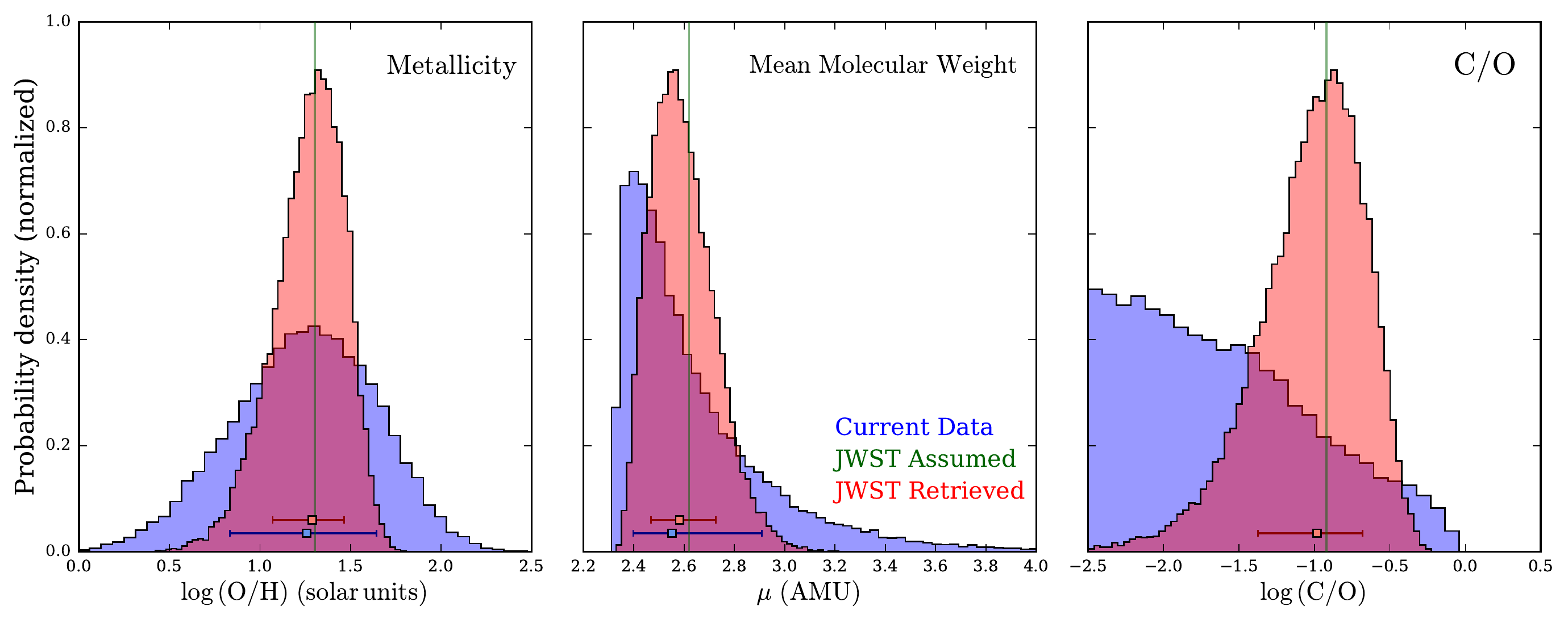}
	\caption{\textbf{Derived properties of HAT-P-26b's atmosphere: current inferences vs. JWST predictions}. Posterior distributions for the metallicity, mean molecular weight, and C/O ratio are shown. Inferences from current observations (blue) are compared against predicted constraints from retrieving synthetic JWST GTO data of HAT-P-26b (red). The chosen values used to generate the synthetic JWST observations are shown by green vertical lines.   Metallicity and C/O are given on a logarithmic scale. `AMU' denotes atomic mass units. Median derived values and 1$\sigma$ confidence levels are denoted by blue and red error bars, for the current and simulated JWST retrievals, respectively. No error bar is given for C/O for current observations, as only an upper limit can be placed (see figure \ref{fig:derived}).}
    \label{fig:JWST_derived}
\end{figure*}

\subsubsection{Metallicity, mean molecular weight, and C/O} \label{subsubsection:JWST_results_derived}

Finally, we examine the precision of derived atmospheric properties of HAT-P-26b obtainable with JWST. In Figure \ref{fig:JWST_derived}, we compare the metallicity, mean molecular weight, and C/O retrieved from current observations (section \ref{subsubsection:derived}) with our predicted constraints from simulated JWST GTO data. We see an improvement in metallicity and mean molecular weight determinations by a factor of $\sim$ 2 (0.4 $\rightarrow$ 0.2 dex and 0.26 $\rightarrow$ 0.13, respectively). C/O changes more dramatically, with the upper limit transformed into a bounded constraint $\sim$ 0.35 dex wide (due to detections of CH$_4$, CO$_2$, and CO). All three derived properties match the values used to generate the synthetic observations to well within 1$\sigma$. We have thus demonstrated the ability of retrievals without the assumption of chemical equilibrium to produce unbiased and precise constraints on metallicities and C/O ratios.

\section{Summary And Discussion} \label{section:discussion}

We have conducted an extensive atmospheric retrieval analysis to reveal the atmospheric composition of the exo-Neptune HAT-P-26b. By simultaneously retrieving space and ground-based transmission spectra, this work represents the most precise determination of atmospheric properties for an extrasolar ice-giant to date. Our major findings are as follows:

\begin{enumerate}
    \item We confirm the presence of H$_2$O in HAT-P-26b's atmosphere (7.2$\sigma$ confidence). The H$_2$O abundance is precisely constrained to $1.5^{+2.1}_{-0.9}\%$ (a precision of 0.4 dex).
    \item We revise the metallicity to $18.1^{+25.9}_{-11.3} \, \times$ solar. This metallicity is robustly super-solar ($>$ 3$\sigma$) and consistent with the solar system mass-metallicity trend to 2$\sigma$.
    \item Non-detections of carbon-bearing species place an upper bound of C/O $<$ 0.33 (to 2$\sigma$).
    \item Features in the optical transmission spectrum provide evidence for the presence of metal hydrides with a combined significance of 4.1$\sigma$. Three potential candidate species are identified: TiH (3.6$\sigma$), CrH (2.1$\sigma$), and ScH (1.8$\sigma$).
    \item The temperature at the terminator is $T_{1 \, \mathrm{mbar}} = 563^{+59}_{-55}$ K, with a temperature gradient of $\sim$80 K across the pressure range probed by the observations. This shallow gradient allows isothermal temperature profiles to explain current observations.
    \item The presence of clouds or hazes are not statistically supported by the present observations. Substructure in the optical and near-infrared spectral regions is better explained by the presence of metal hydrides.
\end{enumerate}
    
Following our analysis of current observations, we additionally conducted retrievals on synthetic JWST observations. We thereby make the following predictions for the JWST GTO campaigns targeting HAT-P-26b: 

\begin{enumerate}
\item Carbon-bearing species will be detectable with JWST. Even at the low C/O established by present observations, CO$_2$, CH$_4$, and CO will be detectable at estimated significance levels of 13.1$\sigma$, 6.2$\sigma$, and 3.7$\sigma$, respectively.
\item A conclusive detection of metal hydrides ($>$ 13$\sigma$) is achievable, including identification of several specific species at $>$ 5$\sigma$ confidence. The metal hydride hypothesis will be directly testable by NIRISS SOSS observations.
\item Abundances for all species detected at $>$ 5$\sigma$ confidence can be constrained to precisions $<$ 0.3 dex. CO constraints are marginally weaker, at $\approx$ 0.5 dex.
\item The H$_2$O abundance, and hence metallicity, can be determined to 0.2 dex precision. Metallicity and mean molecular weight constraints will improve by a factor of $>$ 2 compared to current inferences.
\item C/O will be constrainable to $<$ 0.4 dex.
\item Temperature constraints will improve to $\pm$ 20 K.
\item Abundances are biased by $\sim$ 0.1-0.2 dex under the assumption of an isothermal atmosphere, when in reality a shallow ($\lesssim$ 100~K) temperature gradient is present.
\end{enumerate}

\noindent We now turn to discuss the implications of these findings.

\subsection{HAT-P-26b: an ice giant in context} \label{subsection:mass-metallicity}

Our abundance constraints revise the previous suggestion of a low atmospheric metallicity \citep{Wakeford2017} for HAT-P-26b. Though our values are consistent to within 1$\sigma$, the enhanced precision afforded by our inclusion of ground-based transmission spectra rules out a solar metallicity to $>$ 3$\sigma$. There are, however, some key differences between our retrieval analyses. First, we consider possible contributions of gas phase metal oxides and hydrides as free parameters, both of which provide optical and near-infrared opacity. The analysis of \citet{Wakeford2017} did not include these species, forcing their retrievals to include a cloud deck to provide optical opacity. This cloud deck, in turn, truncates the amplitude of the H$_2$O feature at 1.4$\micron$, reducing the quality of the fit and potentially resulting in an underestimation of the H$_2$O abundance. Our retrieval analysis, whilst including the possibility of clouds, does not find them necessary to explain the present observations. Secondly, the metallicity of \citet{Wakeford2017} was derived by a model averaging procedure, whereby the inclusion of models assuming chemical equilibrium contributed disproportionately, despite not improving the overall fit, due to the lower number of free parameters. By contrast, we are able to explain the full peak-to-trough H$_2$O feature amplitude, without assuming chemical equilibrium (and with as few as 7 free parameters), and thereby obtain an accurate constraint on the H$_2$O abundance and other atmospheric properties.

\begin{figure}
	\includegraphics[width=\linewidth,  trim={0.4cm 1.4cm 0.0cm 0.4cm}]{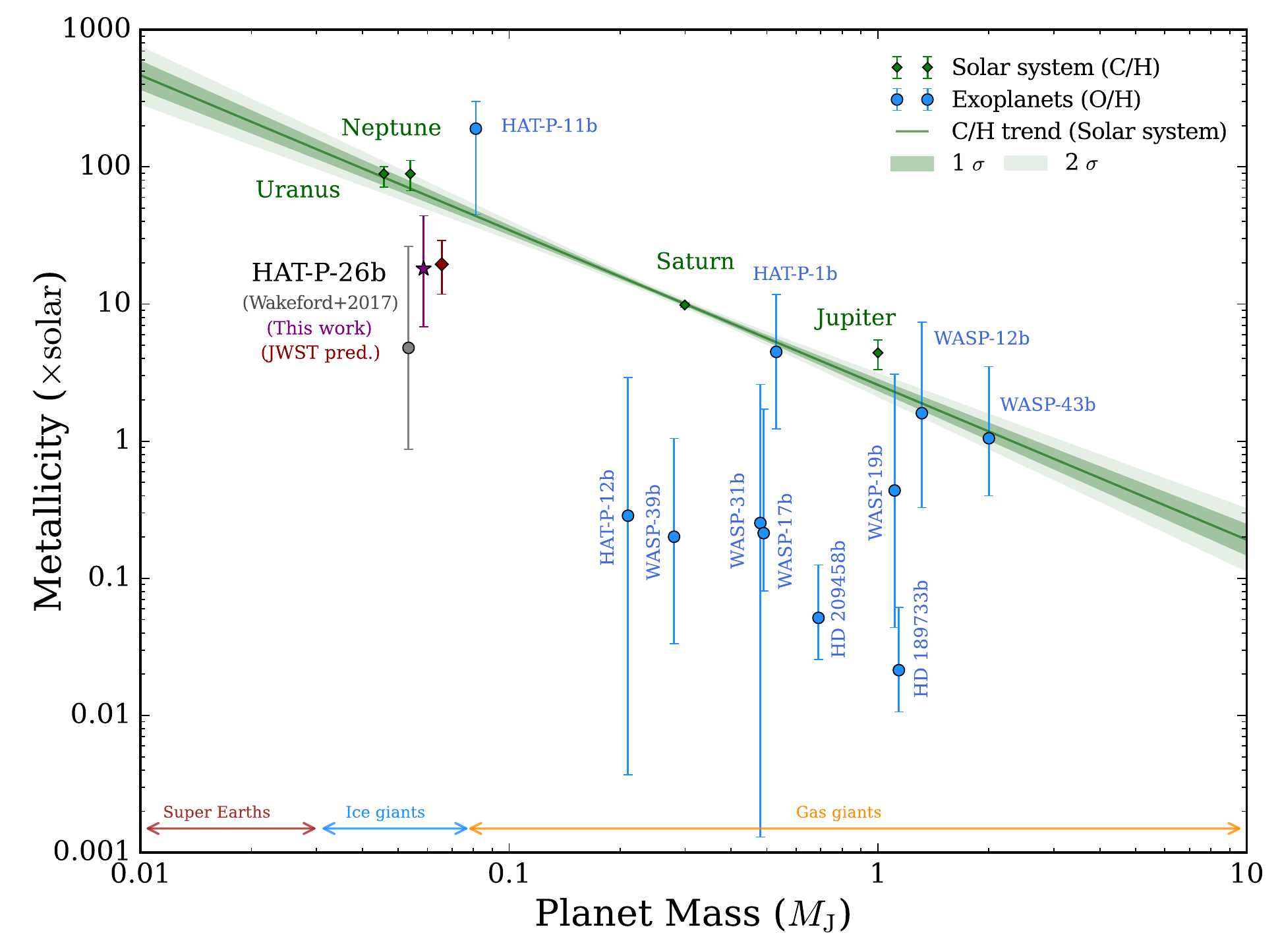}
	\caption{\textbf{Mass-metallicity diagram for exoplanets and the solar system giant planets}. Metallicity refers to $\mathrm{\frac{M/H}{M/H_{\odot}}}$, where `M' is any element heavier than He (see section \ref{subsubsection:derived} for a discussion). Here, M/H is taken as C/H for the solar system giant planets (determined via CH$_4$ abundances), whilst for exoplanets the O/H ratio is taken instead (determined via H$_2$O abundances). The metallicities for Jupiter \citep{Atreya2016}, Saturn \citep{Atreya2016}, Neptune \citep{Karkoschka2011}, and Uranus \citep{Sromovsky2011} (green error bars), follow a linear trend in $\mathrm{log}(M_p) - \mathrm{log}(\mathrm{C/H})$ space. The median linear fit to the solar system metallicities is shown by the green line with corresponding 1$\sigma$ and 2$\sigma$ confidence intervals. Exoplanets with metallicities derived from transmission spectra with detected H$_2$O features are overplotted \citep{Kreidberg2014,Fraine2014,Sedaghati2017,Pinhas2019} (blue error bars). The previously reported metallicity of HAT-P-26b \citep{Wakeford2017} (grey circle) is compared against our revised value (purple star) and our predicted metallicity determination for the JWST GTO campaigns (red diamond). The previous metallicity and JWST prediction have been displaced left and right, respectively, for clarity.}
	\vspace{-0.4cm}
    \label{fig:mass_metallicity_diagram}
\end{figure}

Precise measurements of H$_2$O abundances in exo-Neptune atmospheres provide a fundamental component of ice giant compositions missing for the solar system ice giants, Uranus and Neptune. In particular, the H$_2$O abundance serves as a good indicator for the atmospheric oxygen abundance. At HAT-P-26b's inferred atmospheric temperature ($\approx 560$ K), most of its oxygen is expected to be in H$_2$O, irrespective of the C/O ratio \citep{Madhusudhan2016}. Our precise inference of an O/H ratio of $18.1^{+25.9}_{-11.3} \times$ solar therefore places a strong constraint on the oxygen abundance of an ice giant atmosphere. On the other hand, the H$_2$O abundance, and hence the oxygen abundance, is not known for the solar system ice giants. This is due to the low temperatures of Uranus and Neptune ($<$ 50 K), which causes all the H$_2$O to be sequestered in clouds sufficiently deep in the atmosphere to be inaccessible to observations \citep{Atreya2016}. Only their carbon abundance is known, via CH$_4$ abundance inferences, yielding C/H values of $80 \pm 20 \times$ solar \citep{Sromovsky2011,Karkoschka2011}.

We compare our revised O/H ratio for HAT-P-26b to both the solar system C/H trend and inferred O/H ratios of other exoplanets in Figure \ref{fig:mass_metallicity_diagram}. Note that metallicities derived purely from WFC3 transmission spectra \citep[e.g.][]{Tsiaras2018,Fisher2018} are not shown here, as retrievals without optical data can be highly imprecise due to the `normalisation degeneracy' \citep{Heng2017} and suffer from abundances biased to higher values \citep[see][Fig. 7, for a demonstration]{Pinhas2019}. HAT-P-26b's metallicity lies slightly below the prediction from the solar system mass-metallicity trend, but is consistent to within 2$\sigma$. However, a degree of scatter about this trend is predicted from population synthesis models \citep{Fortney2013}. The precision of HAT-P-26b's metallicity, $\approx$ 0.4 dex, is comparable to the most precisely determined hot Jupiter metallicities \citep[0.35 dex and 0.38 dex for HD 209458b and HD 189733b, respectively, see][]{MacDonald2017a,Pinhas2019}. It is this enhanced precision which robustly establishes the atmosphere of HAT-P-26b as metal enriched.

In contrast, most hot Jupiters appear to be consistent with either solar or sub-solar metallicities \citep{Pinhas2019}. Hot Jupiter metallicity estimates derived from H$_2$O abundances also suffer an additional uncertainty caused by the unknown O content locked into other molecules. For example, at the temperatures of hot Jupiters, around half the O is expected to reside in species other than H$_2$O, especially CO, for a solar C/O ratio \citep{Madhusudhan2012}. Many of the hot Jupiter metallicities in Figure \ref{fig:mass_metallicity_diagram} have therefore either been multiplied by a factor of 2 to account for this or the H$_2$O abundance was linked to a metallicity via the assumption of chemical equilibrium in their respective studies \citep{Kreidberg2014,Sedaghati2017,Pinhas2019}. Such ad hoc corrections will not be necessary for analyses of JWST observations, as the missing oxygen content can be directly constrained via the CO abundance (e.g. our predicted JWST CO constraint of $\approx$ 0.5 dex, see section \ref{subsubsection:JWST_results_composition}). However, at the cooler temperature of HAT-P-26b the oxygen content is expected to mostly reside in H$_2$O. Our metallicity determination is therefore unique, representing the first empirically derived giant planet O abundance free from chemical equilibrium and oxygen partition assumptions.

\subsection{Implications for formation conditions} \label{subsection:formation_conditions}

The super-solar H$_2$O abundance we find for HAT-P-26b has important implications for understanding the formation of ice giants. Given their densities requiring significant solid content in the interior, and that H$_2$O is the most dominant ice expected in planetesimals \citep{Mousis2012,Johnson2012}, its atmospheric abundance opens a window into the inventory of planetesimals accreted during the planet's formation. In particular, close-in exo-Neptunes could arise from a number of different formation scenarios. On the one hand, they could have formed in situ in the inner part of the protoplanetary disk and accreted mainly primordial gas with little contamination from oxygen-rich planetesimals during their migration and growth \citep{Rogers2011}. This scenario would lead to atmospheric compositions with nearly solar O/H. On the other hand, they could have formed in the outer regions of the disk and accreted ice-rich planetesimals during their migration and grown \citep{Madhusudhan2014b,Mordasini2016}. This scenario would lead to super-solar O/H and C/H ratios for formation outside of the H$_2$O and CO / CO$_2$ snow lines, respectively, due to the accreted ices. This later scenario is thought to be the case for Uranus and Neptune due to their locations in the outer solar system \citep{Helled2014}.

Our super-solar O/H of $18.1^{+25.9}_{-11.3} \times$ solar suggests HAT-P-26b formed beyond the H$_2$O snow line. Our low C/O ratio ($< 0.33$), and hence low C/H ratio, additionally suggests formation within the CO / CO$_2$ snow lines. The heavy-element content of the inferred metal hydrides could potentially have arisen from solid planetesimals dissolving into the H$_2$/He envelope, as discussed in the next section. Therefore, a consistent scenario is that HAT-P-26b formed between the H$_2$O and CO / CO$_2$ snow lines, accreted mainly oxygen-rich planetesimals, before migrating inwards to its present location.

\vspace{-0.2cm}

\subsection{Plausibility of inferred metal hydrides} \label{subsection:metal_hydride_plausibility}

Our inference of metal hydrides at $>$ 4$\sigma$ confidence is surprising, as gas phase metal hydrides are expected to condense at temperatures above those present at HAT-P-26b's terminator. In higher temperature atmospheres, such as those of L dwarfs, metal hydrides have sufficiently high abundances to impart prominent absorption features at visible wavelengths \citep{Kirkpatrick2005}. Considering however the equilibrium temperature of HAT-P-26b ($T_{\mathrm{eq}}$ = 990 K), the anticipated abundances of TiH and CrH are $\sim 10^{-17}$ and $\sim 10^{-9}$, respectively \citep[ssuming chemical equilibrium and solar elemental abundances, see][]{Woitke2018} -- many orders of magnitude lower than our retrieved abundances. The inference of spectral features attributable to these species is thus at odds with equilibrium expectations.

One potential explanation is an unaccounted instrumental systematic affecting the optical transmission spectra. As we have utilised data taken from the ground \citep{Stevenson2016} and space \citep{Wakeford2017} at different epochs, differing dataset normalisations could arise due to factors such as stellar variability and differing reduction pipelines. Despite this, we saw in Figure \ref{fig:data} that the datasets agree relatively well in the regions of overlapping wavelength coverage. Nevertheless, we allowed a free offset between the datasets in our analysis to account for this possibility. We additionally show in appendix \ref{section:Without_ground_data} that retrievals solely with the space-based observations still indicate the presence of metal hydrides (albeit at a reduced significance of 2.4$\sigma$). The agreement between the ground-based and space-based observations suggests that the observed optical substructure is not of instrumental origin, and strengthens the interpretation of an atmospheric origin for these features.

The evidence for metal hydrides requires an extreme violation of local chemical equilibrium. One manner to achieve disequilibrium chemistry is vertical transport of material from the deep atmosphere, where the local conditions exceed the condensation temperature. However, whilst vertical mixing can enhance some molecular abundances by $\sim$ 2-3 orders of magnitude over equilibrium expectations \citep{Moses2011,Venot2012}, such mechanisms would struggle to explain our retrieved metal hydride abundances (Figure \ref{fig:abundances}).

We therefore propose an alternative scenario: secular contamination of the atmosphere by metal-rich planetesimals \citep[e.g.][]{Turrini2015,Pinhas2016}. In this scenario, solid planetesimals rich in minerals containing heavy elements (Ti, Fe, Cr, etc.) impact the planet. The resulting high temperatures dissocite chemical bonds, enriching the atmosphere in these metals during the destructive breakup of the impactor. In our case, high-temperature chemistry in the shocks and fireballs following an impact result in new molecules forming \citep{Borunov1997}, potentially producing transient signatures of metal hydrides.

The solar system provides many examples of this process. A well-known example is the impact of comet Shoemaker-Levy 9 into Jupiter, whereafter metallic species such as Fe and Cr were observed \citep{Crovisier1995}. Notable CO abundances in the stratospheres of Neptune and Saturn have been attributed to a $\sim$ km-sized cometary impact within the last 200 years \citep[e.g.][]{Lellouch2005,Cavali2010,Moses2017}. Recently, Mg+ ions in the upper atmosphere of Mars have also been attributed to ablation of metallic meteorites \citep{Crismani2017}.

We examine the plausibility of this scenario via a simple calculation. The calculation is conducted for TiH and CrH, the metal hydrides with the highest significances (3.6$\sigma$ and 2.1$\sigma$, respectively), with the plausibility of ScH discussed separately. Consider a planetesimal with a mass representative of solar system asteroids: $m_{\mathrm{impact}} \approx 10^{18}$ kg \citep{Carry2012}. Taking this planetesimal as analogous to M-type asteroids, its composition may be estimated by noting that spectroscopic studies have suggested these objects may be the progenitors of iron meteorites \citep[e.g.][]{Fornasier2010}. Such meteorites, in turn, tend to contain $>$ 90\% Fe by mass \citep{Buchwald1977}, along with minerals containing Ti (e.g. perovskite, CaTiO$_3$, ilmenite, FeTiO$_3$) and Cr (e.g. chromite, FeCr$_2$O$_4$) \citep{Burbine2016}. We estimate the Ti and Cr mass fraction of the asteroid by taking solar Ti/Fe and Cr/Fe ratios ($\sim$ 3 $\times 10^{-3}$ and $\sim 10^{-2}$, respectively \citet{Asplund2009}). The masses contributed by a single impactor are thus $m_{\mathrm{Ti, \, \, impact}} \approx 3 \times 10^{15}$ kg and $m_{\mathrm{Cr, \, \, impact}} \approx 9 \times 10^{15}$ kg.

Now consider the atmosphere of HAT-P-26b. We evaluate the masses of Ti and Cr present in the photosphere via $m_{\mathrm{Ti, \, Cr, \, \, atm}} = 4 \pi \int^{r_2}_{r_1} \rho_{\mathrm{Ti, \, Cr}} \, r^2 dr$, where $r_1$ and $r_2$ are the radii where $P = 10^{-1}$ bar and $P = 10^{-4}$ bar (corresponding to the observed pressure range, see Figure \ref{fig:PT}). The Ti or Cr density, $\rho_{\mathrm{Ti, Cr}}$, follows from the 1$\sigma$ abundance ranges (Figure \ref{fig:abundances}). Evaluating this integral, we find $m_{\mathrm{Ti, \, \, atm}} \approx (1-30) \times 10^{14}$~kg and $m_{\mathrm{Cr, \, \, atm}} \approx (1-160) \times 10^{14}$~kg. It is thus possible for a single such impactor to deliver sufficient Ti and Cr to be consistent with the observed TiH and CrH abundances.

However, metal hydrides will not persist in the atmosphere indefinitely, as they will eventually condense and sink into the deep atmosphere. A rough persistence timescale is given by the sedimentation timescale, $\tau_{\mathrm{sed}} \approx H/v_t$, where $H$ is the vertical extent traversed and $v_t$ is the terminal velocity of a condensate particle. Considering a condensate forming at $P = 10^{-2}$ bar, it must fall by $H \approx 2 H_{\mathrm{sc}}$, where $H_{\mathrm{sc}}$ is the scale height, to leave the photosphere ($P \gtrsim 10^{-1}$ bar). We take $v_t = 2 \beta r^2 g (\rho_c - \rho_{\mathrm{atm}}) / 9 \eta$, assuming viscous flow \citep{Ackerman2001}. $\beta$ is the Cunningham slip factor (here $\approx 1$), $r$ is the condensate radius, $\rho_c$ the condensate density (e.g. 4000 kg m$^{-3}$ for CaTiO$_3$), $\rho_{\mathrm{atm}}$ the atmospheric density ($\approx 5 \times 10^{-4}$ kg m$^{-3}$ at $P = 10^{-2}$ bar), and $\eta$ the dynamic viscosity ($\approx 10^{-5}$ Pa s for H$_2$ at $T \approx 560$ K, \citet{Ackerman2001}). Taking a condensate radius of $r \approx 1 \micron$, we estimate $v_t \approx 0.01$ ms$^{-1}$ and hence $\tau_{\mathrm{sed}} \approx 10^{8}$~s $\approx 4$ yr. If the observed signatures are to persist over time, replenishment rates of $\dot{m}_{\mathrm{Ti, \, \, atm}} \approx m_{\mathrm{Ti, \, \, atm}}/t_{\mathrm{sed}} \approx (3-80) \times 10^{13}$ kg yr$^{-1}$ and $\dot{m}_{\mathrm{Cr, \, \, atm}} \approx (3-400) \times 10^{13}$ kg yr$^{-1}$ would be required. Such rates could be achieved by a single impactor contributing $m_{\mathrm{Ti, \, \, impact}}$ and $m_{\mathrm{Cr, \, \, impact}}$ once every sedimentation timescale (4 yr), or by a greater impact rate of lower-mass asteroids.

The inferred ScH abundance (Figure \ref{fig:abundances}) is likely inconsistent with a common mechanism (the TiH and CrH abundances resemble solar Ti and Cr values). However, the low statistical significance of ScH alone, partly due to a degeneracy with AlO, raises the possibility that an additional `mystery absorber' not included in our models may be at play. Such an absorber would possess prominent infrared features around 1.06, 1.25, 1.6, and 3.9$\micron$ (see Figure \ref{fig:spectral_signatures}). As we have shown that JWST NIRISS can resolve ScH-AlO degeneracies, we further expect such observations to be capable of potentially identifying possible `mystery absorbers'.

\vspace{-0.2cm}

\subsection{Prospects for future observations} \label{subsection:future_observations}

Our predicted JWST inferences compliment previous analyses. Our JWST metallicity precision, 0.2 dex, is higher than the $\approx$ 0.5 dex found by \citet{Greene2016} -- likely due to our inclusion of NIRSpec G395H and NIRISS SOSS 2nd order data. Our C/O precision, 0.35 dex, is less than the $\approx$ 0.2 dex predicted by \citet{Greene2016}. However, our reference C/O (0.12) is lower than the solar or super-solar values used in their retrievals, reducing our precisions by weakening CH$_4$, CO$_2$, and CO absorption. \citet{Schlawin2018} predicted metallicity and C/O precisions for HAT-P-26b $\approx$ 3$\times$ more precise than ours. Their narrower metallicity and C/O posteriors arise from assuming chemical equilibrium. Our analysis does not make this assumption, instead offering conservative constraints accounting for disequilibrium chemistry. \citet{Line2013} posited that metallicities and C/O ratios derived when assuming log-uniform molecular priors are biased, motivating studies to impose chemical equilibrium. Figure \ref{fig:JWST_derived} demonstrates that our approach successfully retrieves the metallicity and C/O without bias. Relaxing the assumption of chemical equilibrium is therefore appropriate for interpreting JWST observations, freeing retrievals to explore the full richness of exoplanetary atmospheres. 

The potential inference of metal hydrides in HAT-P-26b's atmosphere will be directly testable with JWST. We have demonstrated that NIRISS SOSS observations, especially its 2nd order, provide a powerful window to probe heavy metal chemistry in exoplanetary atmospheres. We have shown that a single transit with NIRISS will be sufficient to detect metal hydrides at $>$ 13$\sigma$ confidence, allowing one to confirm, or rule out, their presence. In the interim, our tentative attribution of optical substructure to TiH absorption can be directly tested by searching for the predicted feature at 0.54 $\micron$ (Figure \ref{fig:spectral_signatures}) with Hubble STIS G430 observations. If the presence of metal hydrides is confirmed by future observations, this will represent the first detection of contamination in the atmosphere of an extrasolar planet.

\vspace{-0.2cm}

\section*{Acknowledgements}

R.J.M. acknowledges financial support from the Science and Technology Facilities Council (STFC), UK, towards his doctoral program. We thank Arazi Pinhas for providing metallicity data, Mihkel Kama for an insightful discussion on planetesimal ablation, Natasha Batalha for advice on using PandExo, and the reviewer for providing helpful comments on our manuscript.

\vspace{-0.3cm}

\bibliographystyle{mnras}
\bibliography{HATP26b.bib} 



\appendix

\vspace{-0.6cm}

\section{Retrievals without ground-based observations} \label{section:Without_ground_data}

Here we present the results from a set of retrievals solely on space-based observations (Hubble STIS+WFC3 and Spitzer). This allows an assessment of the sensitivity of our inferences (e.g. chemical detections) to the data sources used. As this reduced dataset is identical to that presented in \citet{Wakeford2017}, a direct comparison can thereby be drawn between our two respective studies. 

We first describe the setup of our retrievals. We consider a reference model with the following chemical species: H$_2$, He, H$_2$O, Na, K, Li, TiO, VO, AlO, CaO, TiH, CrH, FeH, and ScH. We do not include CH$_4$, NH$_3$, HCN, CO, CO$_2$, or C$_2$H$_2$ in these retrievals for two primary reasons: (i) they absorb primarily in the infrared, so removing the LDSS-3C observations (0.7-1.0$\micron$) should not affect inferences therein; and (ii) these species were not detected in our main analysis (section \ref{subsection:results_composition}), and, since the infrared data used here is identical, this conclusion will also hold here. The exclusion of these molecules additionally reduces the number of free parameters to less than than the number of data points ($N_{\mathrm{data}}$ = 27). The P-T profile and cloud/haze parameterisation is identical to the main text. We therefore have a maximum of 23 free parameters for these retrievals.

We initially conducted 16 retrievals to compute detection significances for each chemical species. The Bayesian model comparison, and resulting significances, are given in Table \ref{table:Appendix_composition_models}. We see that the retrievals without ground-based observations produce similar statistical significances to those in the main text, with two key exceptions: (i) the H$_2$O detection significance rises to 9.2$\sigma$, and (ii) TiH is a non-detection without ground-based observations. The former is a consequence of the non-inclusion of CH$_4$ in these retrievals, which thereby over-estimates the detection significance by not accounting for overlap between H$_2$O and CH$_4$ absorption features. The more conservative 7.2$\sigma$ detection in the main text should therefore be taken instead. The latter finding indicates that the evidence for TiH found in the main text arises from the LDSS-3C observations (e.g. Figure \ref{fig:spectral_signatures}). Note however that CrH and ScH are still inferred using only the space-based observations. The finding that the inclusion of ground-based observations strengthens the metal hydride significance demonstrates that both space-based and ground-based observations are consistent with the presence of an optical opacity source attributable to metal hydrides.

\begin{table}
\ra{1.3}
\caption[]{Bayesian model comparison: composition of HAT-P-26b \citep[][data only]{Wakeford2017}}
\begin{tabular*}{\columnwidth}{l@{\extracolsep{\fill}} cccccl@{}}\toprule
$\mathrm{Model}$ & \multicolumn{1}{p{1cm}}{\centering \hspace{-0.4cm} Evidence \\ \centering $ \hspace{-0.2cm} \mathrm{ln}\left(\mathcal{Z}_{i}\right)$}  & \multicolumn{1}{p{1cm}}{\centering Best-fit \\ \centering $ \chi_{r, \mathrm{min}}^{2}$} & \multicolumn{1}{p{1.7cm}}{\centering \hspace{-0.3cm} Bayes \\ \hspace{-0.2cm} Factor \\ \centering $ \hspace{-0.2cm} \mathcal{B}_{0i}$}& \multicolumn{1}{p{1cm}}{\centering \hspace{-0.6cm} Significance \\ \centering \hspace{-0.4cm} of Ref.}\\ \midrule
\textbf{Full Chem} & $ 204.75 $ & $ 4.30 $ & Ref. & Ref.\\
\hspace{0.2 em} No H$_2$+He & $ 184.50 $ & $ 12.9 $ & $ 6.2 \times 10^{8} $ & $6.7 \sigma$ \\
\hspace{0.2 em} No H$_2$O & $ 164.66 $ & $ 9.21 $ & $ 2.6 \times 10^{17} $ & $9.2 \sigma$ \\
\hspace{0.2 em} No Na & $ 205.07 $ & $ 3.47 $ & $ 0.73 $ & N/A \\
\hspace{0.2 em} No K & $ 205.37 $ & $ 3.51 $ & $ 0.54 $ & N/A \\
\hspace{0.2 em} No Li & $ 204.76 $ & $ 3.60 $ & $ 0.99 $ & N/A \\
\hspace{0.2 em} No TiO & $ 205.36 $ & $ 3.56 $ & $ 0.54 $ & N/A \\
\hspace{0.2 em} No VO & $ 205.45 $ & $ 3.55 $ & $ 0.50 $ & N/A \\
\hspace{0.2 em} No AlO & $ 204.98 $ & $ 3.45 $ & $ 0.79 $ & N/A \\
\hspace{0.2 em} No CaO & $ 205.19 $ & $ 3.38 $ & $ 0.64 $ & N/A \\
\hspace{0.2 em} No TiH & $ 205.02 $ & $ 3.70 $ & $ 0.76 $ & N/A \\
\hspace{0.2 em} No CrH & $ 203.45 $ & $ 3.95 $ & $ 3.67 $ & $2.2 \sigma$ \\
\hspace{0.2 em} No FeH & $ 205.02 $ & $ 3.56 $ & $ 0.76 $ & N/A \\
\hspace{0.2 em} No ScH & $ 204.02 $ & $ 4.01 $ & $ 2.08 $ & $1.8 \sigma$ \\
\hspace{0.2 em} No M-Oxides & $ 206.74 $ & $ 2.19 $ & $ 0.14 $ & N/A \\
\hspace{0.2 em} No M-Hydrides & $ 203.03 $ & $ 3.12 $ & $ 5.58 $ & $2.4 \sigma$ \\
\bottomrule
\vspace{0.1pt}
\end{tabular*}
$\textbf{Notes}:$ The `Full Chem' reference model includes chemical opacity due to H$_2$, He, H$_2$O, Na, K, Li, TiO, VO, AlO, CaO, TiH, CrH, FeH, and ScH. The `No M-Oxides' model has TiO, VO, AlO, and CaO removed; the `No M-Hydrides' model has TiH, CrH, FeH, and ScH removed. The number of degrees of freedom (d.o.f), given by $N_{\mathrm{data}} - N_{\mathrm{params}}$, is 4 for the reference model ($N_{\mathrm{data}}$ = 27). $\chi_{r, \, \mathrm{min}}^{2}$ is the minimum reduced chi-square ($\chi^2$/d.o.f). The significance indicates the degree of preference for the reference model, highlighted in bold, over each alternative model. N/A indicates no (or negative) evidence ($\mathcal{B}_{\mathrm{ij}} \lesssim 1$) supporting a given chemical species.
\label{table:Appendix_composition_models}
\vspace{-0.2cm}
\end{table}

\begin{table}
\ra{1.3}
\caption[]{Bayesian model comparison: temperature structure and clouds of HAT-P-26b \citep[][data only]{Wakeford2017}}
\begin{tabular*}{\columnwidth}{l@{\extracolsep{\fill}} cccccl@{}}\toprule
$\mathrm{Model}$ & \multicolumn{1}{p{1cm}}{\centering \hspace{-0.4cm} Evidence \\ \centering $ \hspace{-0.2cm} \mathrm{ln}\left(\mathcal{Z}_{i}\right)$}  & \multicolumn{1}{p{1cm}}{\centering Best-fit \\ \centering $ \chi_{r, \mathrm{min}}^{2}$} & \multicolumn{1}{p{1.7cm}}{\centering \hspace{-0.3cm} Bayes \\ \hspace{-0.2cm} Factor \\ \centering $ \hspace{-0.2cm} \mathcal{B}_{0i}$}& \multicolumn{1}{p{1cm}}{\centering \hspace{-0.6cm} Significance \\ \centering \hspace{-0.4cm} of Ref.}\\ \midrule
\textbf{P-T + Clouds} & $ 207.93 $ & $ 1.50 $ & Ref. & Ref.\\
\hspace{0.2 em} No Haze & $ 208.10 $ & $ 1.26 $ & $ 0.84 $ & N/A \\
\hspace{0.2 em} Clear Skies & $ 208.14 $ & $ 1.11 $ & $ 0.81 $ & N/A \\
\midrule
\textbf{Iso + Clouds} & $ 208.20 $ & $ 1.05 $ & Ref. & Ref.\\
\hspace{0.2 em} No Haze & $ 208.30 $ & $ 0.93 $ & $ 0.90 $ & N/A \\
\hspace{0.2 em} Clear Skies & $ 208.19 $ & $ 0.85 $ & $ 1.00 $ & N/A \\
\bottomrule
\vspace{0.1pt}
\end{tabular*}
$\textbf{Notes}:$ Two reference models are considered: (i) `P-T' models with a pressure-temperature profile; (ii) `Iso' models with an isothermal atmosphere. The reference models include an opaque cloud deck and a uniform-with-altitude haze. All models include opacity due to H$_2$, He, H$_2$O, TiH, CrH, and ScH. The number of degrees of freedom (d.o.f), given by $N_{\mathrm{data}} - N_{\mathrm{params}}$, is 12 for the `P-T' reference model and 17 for the `Iso' reference model ($N_{\mathrm{data}}$ = 27). $\chi_{r, \, \mathrm{min}}^{2}$ is the minimum reduced chi-square ($\chi^2$/d.o.f). The significance indicates the degree of preference for a reference model, highlighted in bold, over each alternative model. N/A indicates no (or negative) evidence ($\mathcal{B}_{\mathrm{ij}} \lesssim 1$) supporting hazes or clouds.
\label{table:Appendix_PT_cloud_models}
\end{table}

The posterior distribution of the `full chem' model, along with accompanying parameter constraints, is provided in the supplementary online material. The inferences are consistent with those in the main text, though constraints are less precise without ground-based observations. For example, the H$_2$O abundance precision is $\approx$ 0.5 dex here, compared to $\approx$ 0.4 dex when including ground-based data. Comparing with the H$_2$O abundance and temperature derived by \citet{Wakeford2017} using a similar retrieval approach (log($X_{\mathrm{H_2 O}}$) $\sim -2.2 \pm 0.7$ and $T \sim 550^{+150}_{-100}$ K, their Figure S4), our values are seen to be consistent.

Finally, we performed a complexity analysis to assess the necessity of non-isothermal temperature structures and clouds/hazes. This analysis, identical to that in section \ref{subsection:results_model_complexity}, is represented in Table \ref{table:Appendix_PT_cloud_models}. We find no evidence supporting models including clouds or hazes, with the Bayesian evidence expressing a slight preference towards isothermal models. As in the main text, isothermal, clear, models with H$_2$O and metal hydride opacity provide the optimum minimal complexity model ($\chi_{r, \mathrm{min}}^{2} \approx 1$). This is the most notable difference between our results and those of \citet{Wakeford2017}, where a grey cloud deck was invoked to explain the optical data. We have run additional retrievals without metal hydride or metal oxide opacity to verify that we reproduce their inference of a cloud deck when making commensurate assumptions, though the Bayesian evidence of such models is low ($\approx 201.3$). We therefore conclude that current space-based observations offer no statistically significant evidence for the presence of clouds in HAT-P-26b's atmosphere.

\vspace{-0.2cm}


\bsp	
\label{lastpage}
\end{document}